\documentclass[aps,preprint,groupedaddress]{revtex4}  

\usepackage{amssymb}
\usepackage{amsmath}
\usepackage{amssymb}
\usepackage{graphicx}
\usepackage{subfigure}
\usepackage{color}
\usepackage{mathrsfs}
\usepackage[dvipsnames]{xcolor}

\usepackage[breaklinks=true,colorlinks=true]{hyperref}
\hypersetup{colorlinks=true,citecolor=blue,linkcolor=blue,urlcolor=blue}

\usepackage[utf8]{inputenc}

\begin{document}


\title{Scattering of kinks in scalar-field models \\with higher-order self-interactions}


\author{Aliakbar Moradi Marjaneh$^{1}$, Fabiano C. Simas$^{2,3}$ and D. Bazeia$^{4}$}

\email{moradimarjaneh@gmail.com; fc.simas@ufma.br; bazeia@fisica.ufpb.br}

\affiliation{$^{1}$Department of Physics, Quchan Branch, Islamic Azad University, Quchan, Iran\\
$^{2}$Programa de P\'os-Gradua\c c\~ao em F\'isica, Universidade Federal do Maranh\~ao
(UFMA), Campus Universit\'ario do Bacanga, 65085-580, S\~ao Lu\'is, Maranh\~ao, Brazil\\
$^{3}$Departamento de F\'isica, Universidade Federal do Maranh\~ao (UFMA), Campus Universit\'ario do Bacanga, 65085-580, S\~ao Lu\'is, Maranh\~ao, Brazil\\
{$^{4}$Departamento de F\'isica, Universidade Federal da Para\'iba (UFPB), 58051-970 Jo\~ao Pessoa, PB, Brazil}}


\begin{abstract}

Higher-order scalar field models in two dimensions, including the $\phi^8$ model, have been researched. It has been shown that for some special cases of the minima positions of the potential, the explicit kink solutions can be found. However, in physical applications, it is very important to know all the explicit solutions of a model for any minima position. In the present study, with the help of some deformation functions, we have shown that higher-order scalar field theories can be obtained with explicit kinks. In particular, we introduced two deformation functions that, when applied to the well known $\phi^4$ and $\phi^6$ models, produce modified $\phi^8$ and $\phi^{10}$ models, respectively, with all their explicit kink-like solutions which depend on a single parameter. Since this parameter controls the position of the minima of the potential, we have found interesting new solutions in many distinct cases. We have also studied the kink mass, the behavior of the excitation spectra and several kink-antikink collisions for these two new modified models. The collision outcome is determined by the initial configuration, specifically the sequence in which the kink-antikink and antikink-kink pairings emerge. Another interesting finding is the suppression of resonance windows, which may be explained by the presence of a set of internal modes in the model.

\end{abstract}




\maketitle


\section{Introduction}
\label{sec:introduction}


Topological configurations such as stable particle-like objects with smooth structure and finite mass, may play important role in physics. They are solutions of field-theoretical models that have been used for at least sixty years on different scales \cite{Ginzburg.ZhETF.1950,Bishop.PhysD.1980,Rajaraman.book.1982, Coleman.book.1985,Vilenkin.book.2000,Manton.book.2004,Dauxois.book.2006, Vachaspati.book.2010,Kevrekidis.book.2019, Optics1}, and are of interest to cosmology, high energy physics, condensed matter physics and optics. Scientists have always tried to introduce models that can describe the physical behavior of real examples, and they usually start with simple models in $(1+1)$ dimensional space-time and gradually add new degrees of freedom and dimensions to achieve more complex situations. Among the several possibilities, famous models such as the integrable sine-Gordon system \cite{Rajaraman.book.1982, Coleman.book.1985,Vachaspati.book.2010, Javidan.cc.2010, Moradi.EPJB.2018, Mukhopadhyay.JHEP.2022} and non-integrable models such as $\phi^4$ \cite{Kevrekidis.book.2019,mainak2023, Campbell.PhysD.1983,anninos.1991, Belova.PhysD.1988, Goodman.SIAM_JADS.2005, Moradi.CNSNS.2017, Dorey.JHEP.2017, Dorey.PLB.2018, Weigel.JPCS.2014, Weigel.PRD.2016,  Askari.CSF.2020} and $\phi^6$ \cite{Weigel.JPCS.2014, Weigel.PRD.2016, Dorey.PRL.2011, Gani.PRD.2014, Romanczukiewicz.PLB.2017, Weigel.PLB.2017, Weigel.AHEP.2017, Moradi.JHEP.2017, Demirkaya.JHEP.2017, Lima.JHEP.2019, Moradi.EPJB.2022,Saadatmand.2023} have been investigated in the field of nonlinear physics. By deforming these models or creating different conditions, more complex models are obtained, which are usually closer to reality. The models obtained by changing the sine-Gordon model, such as the double sine-Gordon model, belong to this category \cite{Peyrard.PhysD.1983.msG, Campbell.PhysD.1986.dsG, Bazeia.EPJC.2011, Bazeia.EPJC.2013, Gani.EPJC.2018, Belendryasova.JPCS.2019, Gani.EPJC.2019,riazi}. Different models, whether periodic or polynomial, are obtained by {\it deformation procedure}, which have applications in different fields \cite{Bazeia.PRD.2002, Bazeia.PRD.2004, Bazeia:2005hu, Bazeia.PRD.2006, Bazeia.EPJC.2018, Moradi.CSF.2022}. For example, black holes, tachyon matter cosmology and quintessential inflation can be studied with hyperbolic models \cite{QiangWen.PRD.2015, Pourhassan.IJMPD.2017, Agarwal.PLB.2017}. Although more complex models are richer in explaining the physical behavior of various examples, it becomes more difficult to obtain analytical solutions for them. For example, in solving the nonlinear Klein-Gordon equation for potentials of order eight and higher, there are analytical solutions just for some cases where the minima of the potential are located in specific positions; see, e.g., Refs.  \cite{Bazeia.PRD.2006,Gani.PRD.2020, Khare.2021}.

Kink scattering in many models has revealed the complexities of the outcomes. In particular, in Ref. \cite{Campbell.PhysD.1983} the study shows the kink-antikink scattering in the $\phi^4$ model. The explanation for the appearance of the two-bounce windows is related to the presence of the shape mode. In another important work \cite{Dorey.PRL.2011}, the formation of resonant windows was analyzed, however, the explanation was based on the presence of vibrational states due to a perturbation potential for the antikink-kink pair. Another counterexample to the mechanism of energy exchange between translational and vibrational modes can be seen in Ref. \cite{simas.2016}. Notably, two-bounce windows are suppressed even when internal modes are present. Other works that address the collision of kinks and depict various types of outcomes can also be cited. For instance, the scattering between wobbling kinks \cite{alonso.2021,joao.2021,joao.2021.2}, investigation of models with more scalar fields \cite{alonso.2020,alonso.2018,halava.2012} and a kink-antikink scattering scenario with an additional scalar field located in its quantum vacuum \cite{vacha.2023}. In addition, it is worth noting that kink scattering may appear under the presence of a spectral wall, giving rise to interesting new results \cite{adam.2019,joao.2023,adam.2022,adam.2021}. Moreover, a recent investigation has examined how a deformation function in scalar field models can change the solution, mass, internal structure and collision process of kinks, also leading to new results of current interest  \cite{Moradi.CSF.2022}.

Motivated by the above new results, in this work we study properties of two polynomial potentials, the deformed $\phi^8$ and $\phi^{10}$ models, described by deformations of the standard $\phi^4$ and  $\phi^6$ models. Specifically, higher order self-interaction models are capable of simulating massless mesons, which can result in long-range interactions \cite{lohe}. Furthermore, a second-order phase transition followed by a first-order one may be used to characterize the $\phi^8$ model \cite{told1,told2,saxena3}. In condensed matter, for example, it was investigated the elastic characteristics of crystals, whose results are connected to successive first-order phase transitions \cite{mroz}, which can also be applicable to isomorphous phase transitions \cite{liber} and that may occur in materials science \cite{gufan,saxena2}. In circumstances of consecutive phase transitions, it is of interest to include a $\phi^{10}$ field theory.  In particular, it is worth noting that this model is related to the study of protein crystallization through the possibility of the formation of cylindrical lattice \cite{boulbi}. Also, a higher-order theory is presented in Ref. \cite{cohen} to give a phenomenology of cubic perovskite ferroelectrics. In this case, the authors first demonstrate that a monoclinic phase cannot be supported by a sixth-order theory, and then they examine a higher-order theory. An additional option is to consider scalar field theories of the type $\phi^{2n}$ as Lane-Emden truncations of a periodic potential \cite{valle}. This would allow these theories to be used as models for dark matter halos. Another important aspect is the investigation of defect collisions using higher-order models. In Ref. \cite{EkaGani}, escape windows were detected in the scattering of $\phi^8$ kinks.

We organize the paper as follows. In section \ref{sec:deformationfunction}, we review the general statement of the deformation procedure. Then, in the next two sections \ref{sec:modifiedphi8model} and \ref{sec:modifiedphi10model}, we define two deformation functions dependent on real parameter $a$, and introduce modified deformed $\phi^8$ and $\phi^{10}$ models. There, we also explore the scattering of kinks. The impact of the internal modes and the distinctions between kink-antikink and antikink-kink collisions are also studied in these two Sections. We also investigate the explicit kinks, their masses, the stability potentials and the values of the internal modes. These two new deformed models, are obtained via a deformation function that depends on a single real parameter, which is used to control the positions of the minima of potentials and can affect the main features of the models and guide us to investigate the two models. The work is closed in Section \ref{sec:conclusion}, where we add our conclusions and comment on issues concerning future directions of new research on the subject.

\section{Deformation procedure}
\label{sec:deformationfunction}

Let us start reviewing the methodology related to the deformation procedure. The dynamics of the real scalar field for a field theoretical model in $(1+1)$ space-time  dimensions is described by the Lagrangian density
\begin{eqnarray}\label{eq:lagrangian}
\mathcal{L}=\frac{1}{2}\left(\frac{\partial \phi}{\partial t}\right)^2-\frac{1}{2}\left(\frac{\partial \phi}{\partial x}\right)^2-V(\phi).
\end{eqnarray}
It yields the Klein-Gordon equation of motion
\begin{equation}\label{eq:EOM}
\frac{\partial^2 \phi}{\partial t^2}  - \frac{\partial^2 \phi}{\partial x^2} + \frac{dV(\phi)}{d\phi}=0.
\end{equation}
In the static case, $\phi=\phi(x)$, this equation can be reduced to the following first-order  equations
\begin{equation}\label{eq:bps}
\frac{d\phi}{dx} = \pm\sqrt{2V(\phi)}.
\end{equation}
We suppose we can solve these equations to find static solutions analytically. In this case, by applying a deformation function, $f(\phi)$, on the potential ${V}(\phi)$ \cite{Bazeia.PRD.2002,Bazeia.PRD.2004,Bazeia.PRD.2006,Bazeia:2005hu,Bazeia.EPJC.2018}, a new deformed model with the deformed potential, $\tilde{V}(\phi)$, is formed
\begin{equation}\label{eq:deformedpotential}
\tilde{V}(\phi)=\frac{V[\phi\to f(\phi)]}{[f^{\prime}(\phi)]^2},
\end{equation}
such that, in the new model the new static kinks are given by 
\begin{equation}\label{eq:deformedkinks}
\Tilde{\phi}_K(x)=f^{-1}[\phi_K(x)].
\end{equation}

For solutions that solve the above first-order equations, we can obtain the kink energy or kink mass as follows
\begin{eqnarray}\label{eq:mass}
mass &=& \int_{-\infty}^{+\infty}\bigg(\frac{\partial\phi}{\partial x}\bigg)^2 dx.  
 \end{eqnarray}
Also, after adding a small perturbation to the static kink, in the form $\phi(x,t)=\phi_K(x)+ \eta(x) \cos (\omega t)$, one can find the Schr\"odinger-like equation which allows studying linear stability and kink excitation spectra
\begin{equation}\label{eq:schrodingerlike}
-\frac{d^2 \eta}{d x^2}+U\eta=\omega^2\eta,
\end{equation}
 where $U=U(x)$ is the kink stability potential, which is obtained from the second derivative of the potential $V(\phi)$ (or $\tilde{V}(\phi)$ for modified model), with respect to the kink $\phi_K(x)$ (or $\tilde{\phi}_K(x)$)

\begin{eqnarray}\label{eq:stabilitypotentials}
U(x)=\frac{d^2V(\phi)}{d\phi^2}\bigg{|}_{\phi=\phi_K(x)}.
\end{eqnarray}
The above Eq. \eqref{eq:schrodingerlike} and the corresponding stability potential are important for the scattering process, since they provide the number of internal modes, which directly influence the output of the collision.

With the above methodology, we can now discuss the modified $\phi^8$ and $\phi^{10}$ models, as well as their respective collision results.


\section{modified  \texorpdfstring{$\phi^8$ }{pdfbookmark} model}
\label{sec:modifiedphi8model}

\begin{figure*}[!ht]
\begin{center}
  \centering
    \subfigure[\quad Deformation function (plus sign, $c_1=0$)]{\includegraphics[width=0.45
 \textwidth]{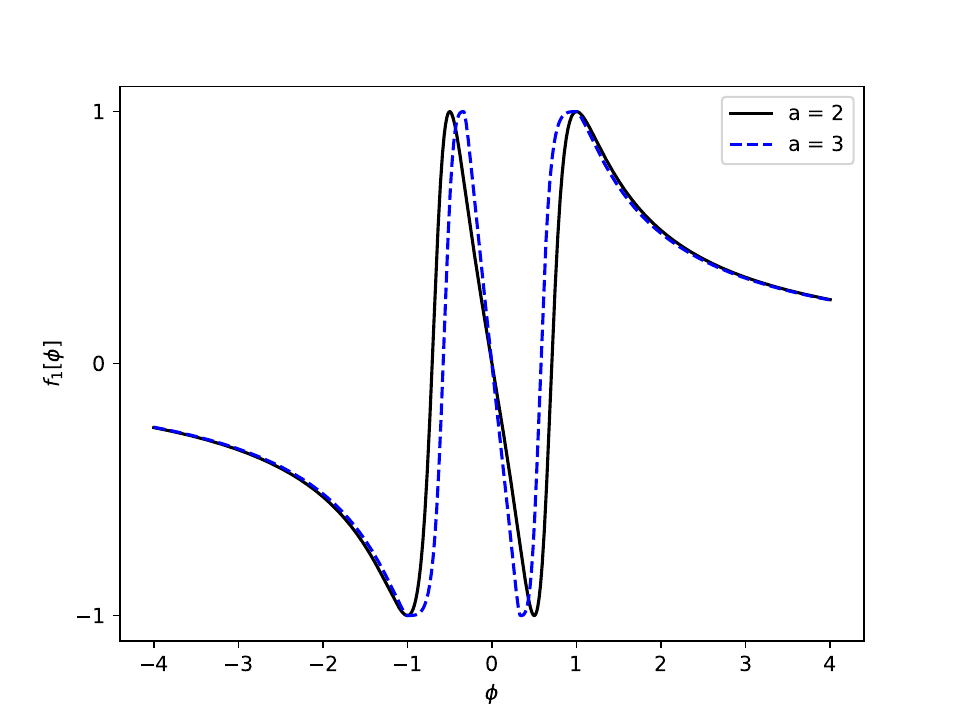}\label{fig:f1}}
  \subfigure[\quad Potentials]{\includegraphics[width=0.45
 \textwidth]{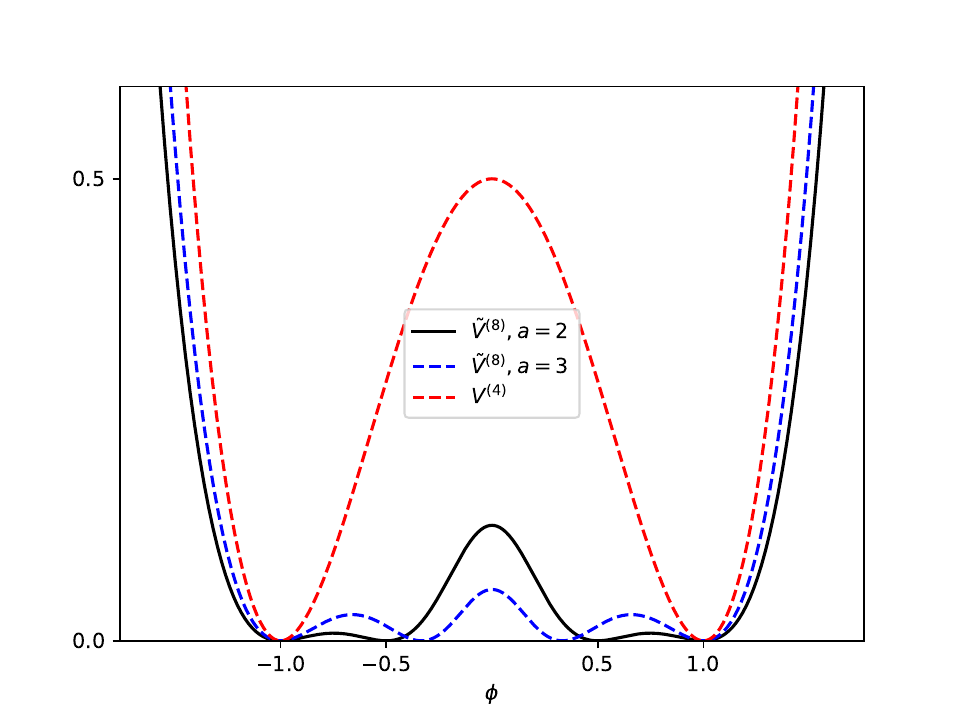}\label{fig:modifiedphi8potentials}}
\\
  \subfigure[\quad Modified $\phi^8$ kinks for $a=2$]{\includegraphics[width=0.45
 \textwidth]{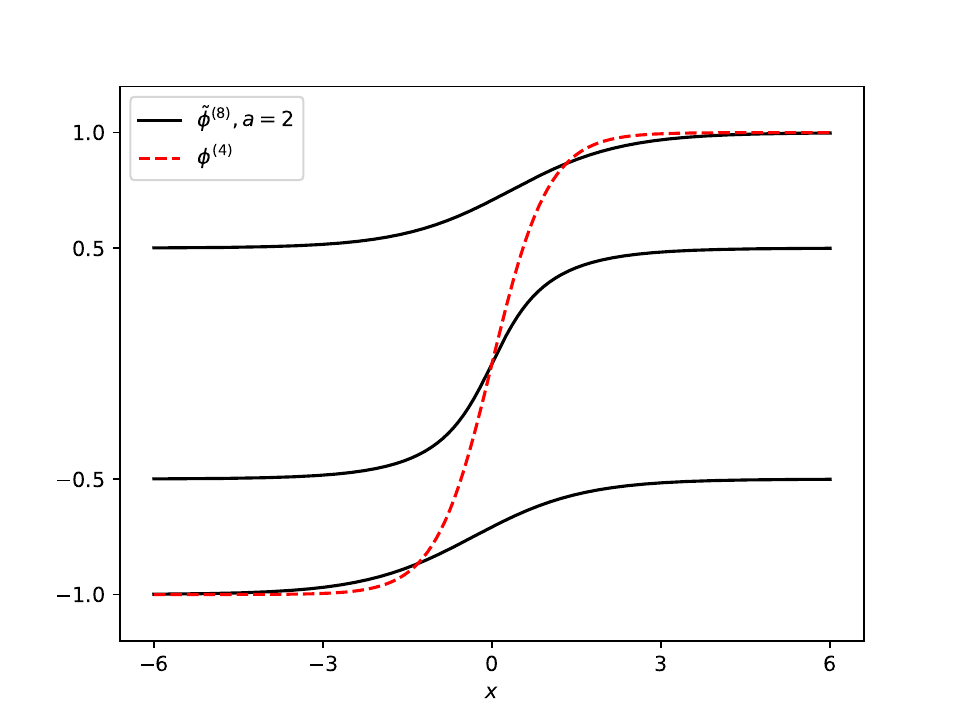}\label{fig:modifiedphi8kinksa2}}
  \subfigure[\quad Modified $\phi^8$ kinks for $a=3$]{\includegraphics[width=0.45
 \textwidth]{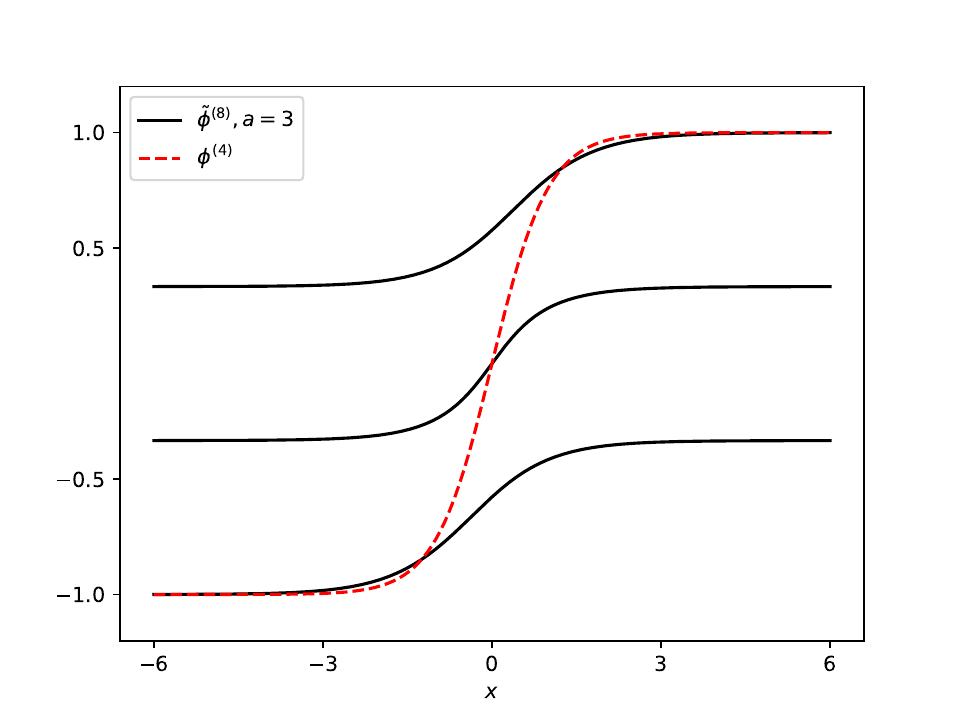}\label{fig:modifiedphi8kinksa3}}
\\
  \caption{(a) The deformation function $f_1[\phi]$ as a function of $\phi$, (b) modified $\phi^{8}$ potential and modified $\phi^{8}$ kinks for (c) $a=2$ and (d) $a=3$. The $\phi^4$ model is represented by the dashed red line.}
  \label{fig:modifiedphi8}
\end{center}
\end{figure*}

First, we introduce the deformation function $f_1[\phi]$, see Fig.~\ref{fig:f1}
\begin{equation}\label{eq:f1}
f_1[\phi]=\pm \tanh \left(\frac{a }{a-1}\left(\tanh ^{-1}(\phi )-\tanh ^{-1}(a \phi )\right)-c_1\right).
\end{equation}
where $a$ is a real parameter and $c_1$ is an arbitrary constant. Now, by using of Eq.~\eqref{eq:deformedpotential}, this deformation function can change the $\phi^4$ model with the potential $V^{(4)}=\frac{1}{2}(1-\phi^2)^2$ to the following modified $\phi^8$ model
\begin{eqnarray}\label{eq:modifiedphi8potential}
\tilde{V}^{(8)}&=& \frac{1}{2}\frac{\left(1-\phi ^2\right)^2 \left(\frac{1}{a^2}-\phi ^2\right)^2}{ \left(\frac{1}{a}+\phi ^2\right)^2}. 
\end{eqnarray}

This potential has four minima at $\phi=\pm 1, \pm \frac{1}{a}$. We consider the parameter equal or greater than one, $a \ge 1$, without changing the generality of the problem. The potential profiles for $a=2$ and $a=3$ are shown in Fig.~\ref{fig:modifiedphi8potentials}. We notice that the point of maximum potential in the central region decreases with increasing $a$. From Eq.~\eqref{eq:deformedkinks}, one can find all explicit kinks solution of the model
\begin{eqnarray}\label{eq:modifiedphi8kinks}
\tilde\phi^{(8)} \to
\begin{cases}
\tilde{\phi}^{(8)}_{AS}(\frac{1}{a},1)=\frac{(1-a) }{2 a}\tanh \left(\frac{(1-a) }{a}x\right)+\sqrt{\left(\frac{1-a}{2 a}\right)^2 \tanh ^2\left(\frac{(1-a) }{a}x\right)+\frac{1}{a}},  \\
\tilde{\phi}^{(8)}_{S}(-\frac{1}{a},\frac{1}{a})=\frac{1}{2 \left(e^{\frac{(2 (a-1)) x}{a}}-1\right)} \Bigg(\left(\frac{1}{a}-1\right) \left(e^{\frac{(2 (a-1)) x}{a}}+1\right)  \\
 \qquad \qquad \qquad  +\sqrt{\frac{4 \left(e^{\frac{(2 (a-1)) x}{a}}-1\right)^2}{a}+\left(\left(1-\frac{1}{a}\right) \left(e^{\frac{(2 (a-1)) x}{a}}+1\right)\right)^2}\Bigg), \\ 
\tilde{\phi}^{(8)}_{AS}(-1,-\frac{1}{a})=\frac{(1-a) }{2 a}\tanh \left(\frac{(1-a) }{a}x\right)-\sqrt{\left(\frac{1-a}{2 a}\right)^2 \tanh ^2\left(\frac{(1-a) }{a}x\right)+\frac{1}{a}}.
\end{cases}
\end{eqnarray}

These kinks include a symmetric kink, $\tilde{\phi}^{(8)}_{S}$, which is in between the minima $-\frac{1}{a}$ and $\frac{1}{a}$, and two similar asymmetric kinks, $\tilde{\phi}^{(8)}_{AS}$, which are located in the sectors $(-1,-\frac{1}{a})$ and $(\frac{1}{a},1)$. As a result of this asymmetry, the behavior of the collision will depend on the initial configuration, since the kink-antikink case differs from the antikink-kink one, yielding different effects. The symmetric and asymmetric solutions are depicted for $a=2$ and $a=3$ in Figs.~\ref{fig:modifiedphi8kinksa2} and \ref{fig:modifiedphi8kinksa3}, respectively. 
 
In the case of $a=1$, the expressions for the potential and kink solution are given by
\begin{eqnarray}\label{eq:modifiedphi8a1}
\begin{cases}
\tilde V^{(8)}(a=1)=\frac{1}{2}\frac{\left(\phi ^2-1\right)^4}{ \left(\phi ^2+1\right)^2},
\\
\tilde{\phi}^{(8)}_{S}(-1,1)=\frac{\sqrt{4 x^2+1}-1}{2 x}.
\end{cases}
\end{eqnarray}

\begin{figure*}[!ht]
\begin{center}
  \centering
    \subfigure[\quad Potential]{\includegraphics[width=0.45
 \textwidth]{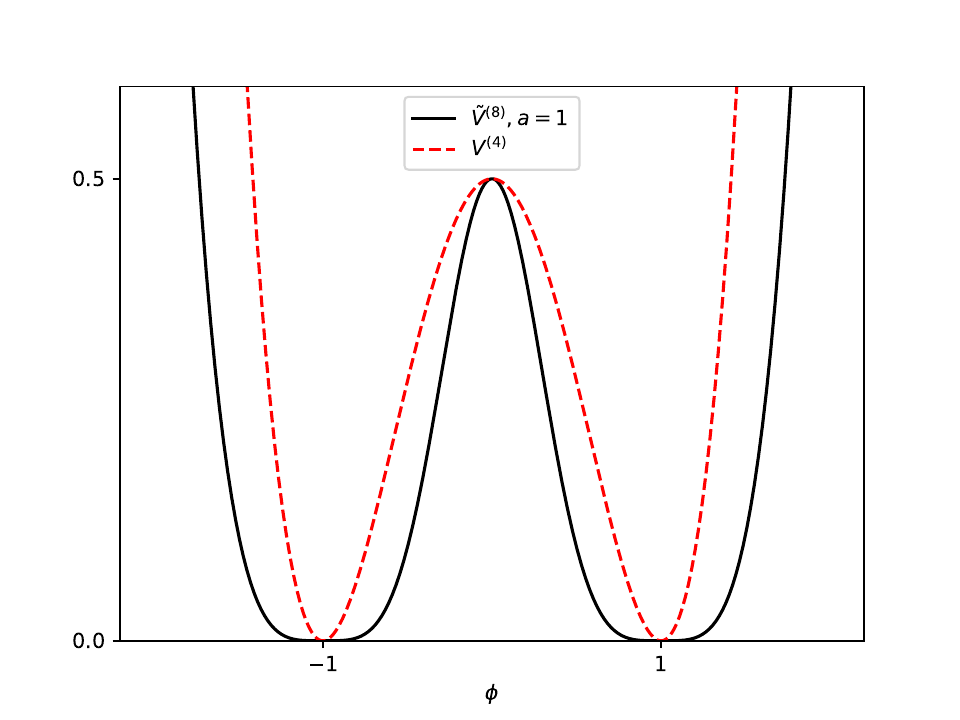}\label{fig:modifiedphi8potentialsa1}}
  \subfigure[\quad Kink]{\includegraphics[width=0.45
 \textwidth]{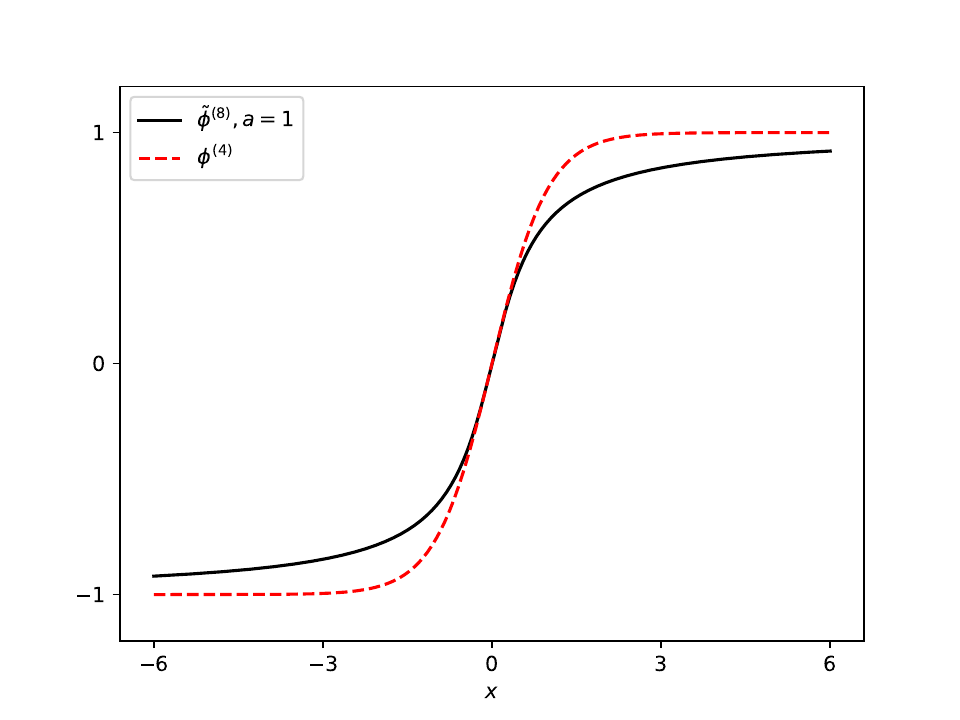}\label{fig:modifiedphi8kinksa1}}
\\
  \caption{Deformed $\phi^8$ potential and kink in the case of $a=1$ compared well known $\phi^4$ model. }
  \label{fig:modifiedphi8a1}
\end{center}
\end{figure*}

In Fig.~\ref{fig:modifiedphi8a1} one depicts the potential and kink solution for $a=1$. It should be emphasized that the potential contains only two minimum at $\pm 1$ and it is interesting to observe that the minima are flatter. In this case, there is just one central kink sector and this behavior results in defects with long-range tails. In Refs. \cite{ivan,diojoaza}, the authors studied in detail the scattering of kinks with long- and short-range tails. In Ref. \cite{ivan}, the results revealed the existence of resonant windows due to the presence of the bound state. In Ref. \cite{diojoaza}, the output showed the decay of the pair in radiation, favoring the appearance of kink-antikink pairs. A novel technique is required for the development of collisions with long-range tails, as described in Refs. \cite{chidede,joazalong}. We shall not handle kink-antikink collisions for the scenario $a=1$. This work proposes to investigate scattering in the presence of both symmetric and asymmetric kinks.
 
When $a \to \infty$, the modified $\phi^8$ potential (Eq.~\eqref{eq:modifiedphi8potential}) tends to well known $\phi^4$ potential, with the symmetric kink solution $\phi^{(4)}=\tanh(x)$, the mass of ${4}/{3}$ and the only shape mode with frequency $\omega_1^2=3$. In this case, the  potential has just two minima, and the asymmetric kinks disappear.

\begin{figure*}[!ht]
\begin{center}
  \centering
    \subfigure[ ]{\includegraphics[width=0.45
 \textwidth]{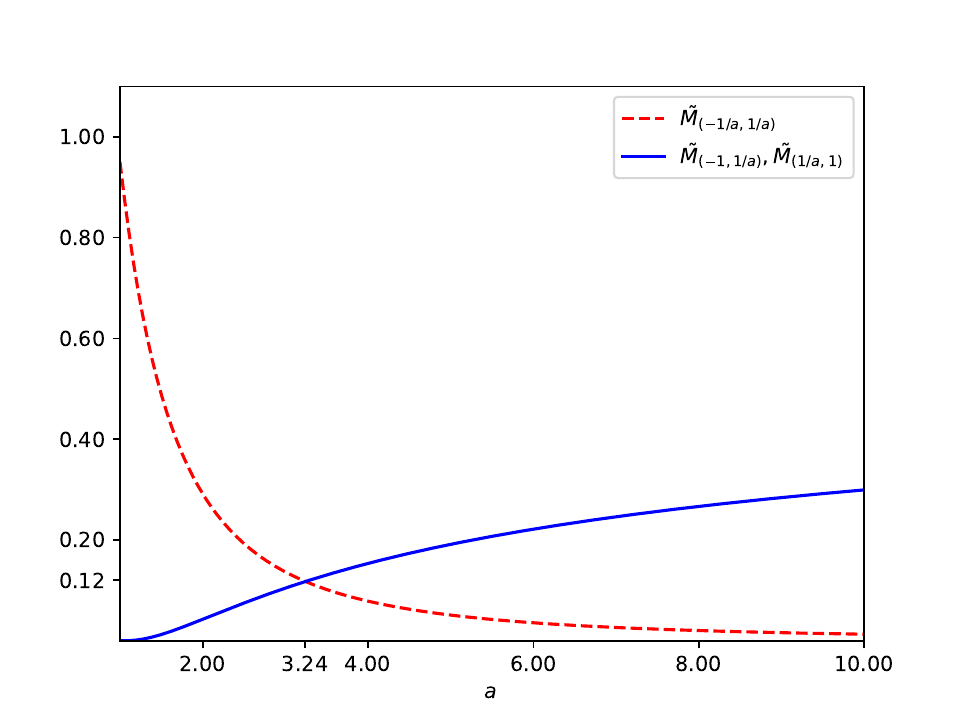}\label{fig:modifiedphi8mass}}
    \subfigure[ ]{\includegraphics[width=0.45
 \textwidth]{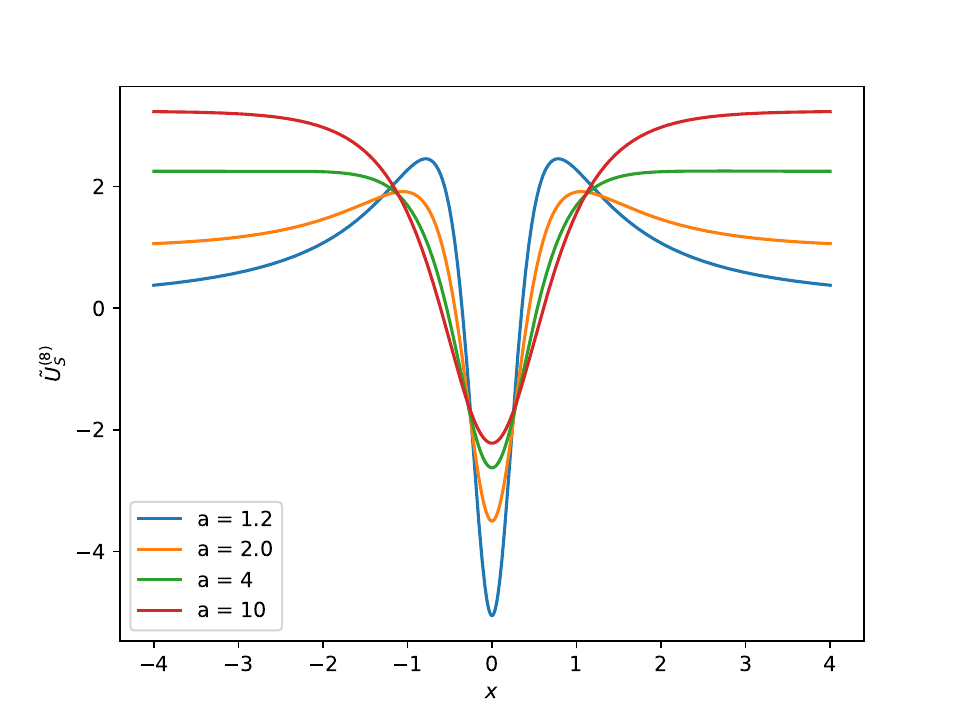}\label{fig:modifiedphi8qmpS}}
\\
  \subfigure[ ]{\includegraphics[width=0.45
 \textwidth]{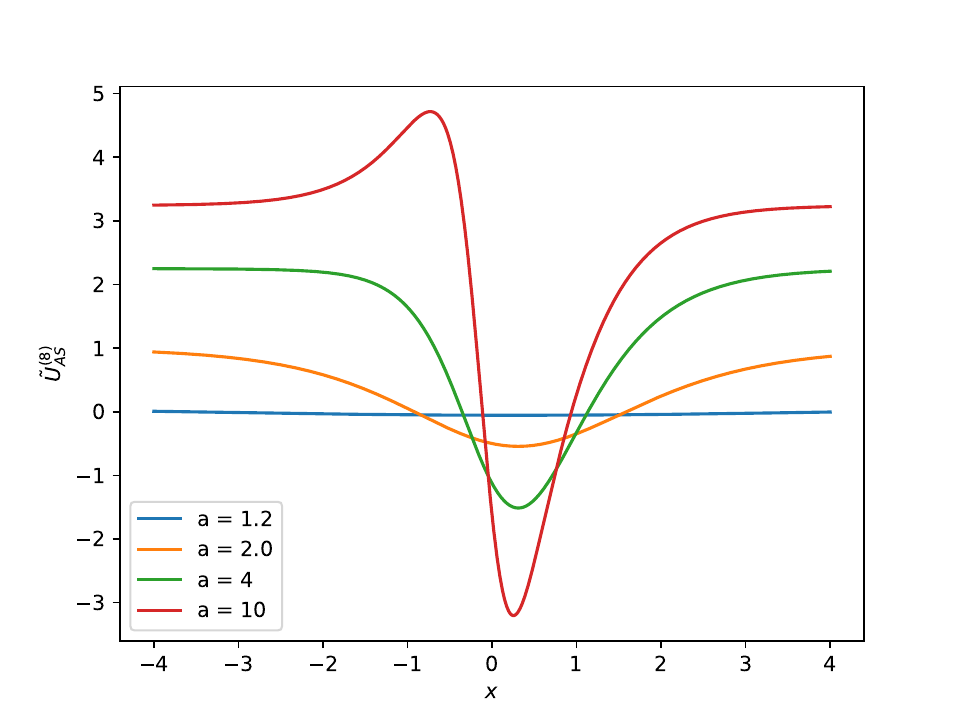}\label{fig:modifiedphi8qmpAS}}
  \subfigure[ ]{\includegraphics[width=0.45
 \textwidth]{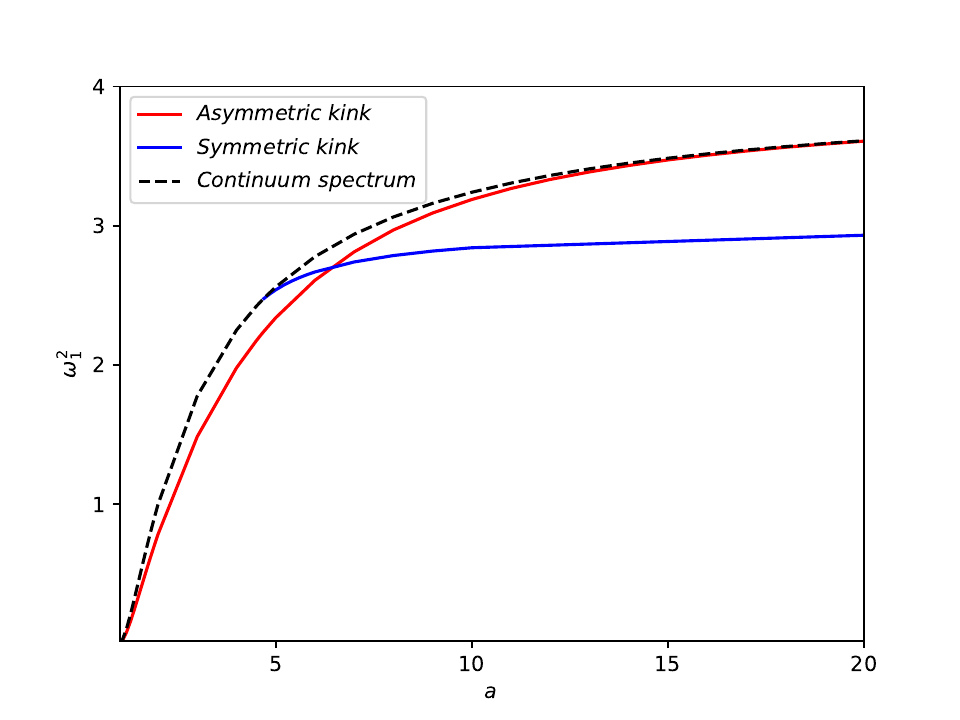}\label{fig:modifiedphi8modes}}
\\
  \caption{Deformed $\phi^{8}$ model: (a) kink mass, (b) and (c) quantum mechanical potential for symmetric and asymmetric kinks, respectively, as a function of $x$ and (d) the squared frequencies $\omega^2_1$ of the vibrational states as a function of the parameter $a$ for symmetric and asymmetric kinks.}
  \label{fig:modifiedphi8massmodesqmp}
\end{center}
\end{figure*}

Different values of parameter $a$ provide the examination of the changes of the kinks relative to each other from the point of view of mass, internal mode and stability potential, as well as the examination of different scatterings. In this general case, the kink masses are obtained in terms of parameter $a$. They are
\begin{eqnarray}\label{eq:modifiedphi8mass}
mass &=& \int_{-\infty}^{+\infty}\bigg(\frac{\partial\phi}{\partial x}\bigg)^2 dx  \to 
\begin{cases}
\tilde M^{(8)}_S=\frac{6 \sqrt{a} (a+1)^2 \tan ^{-1}\left(\frac{a}{\sqrt{a^3}}\right)-6 a (a+1)-4}{3 a^3},
\\
\tilde M^{(8)}_{AS}=-\frac{(a+1)^2 \tan ^{-1}\left(\frac{a-1}{2 \sqrt{a}}\right)}{a^{5/2}}-\frac{2}{3 a^3}+\frac{2}{3},
\end{cases}
\end{eqnarray}
where $\tilde M^{(8)}_S$ is the mass of symmetric kink in sector $(-\frac{1}{a},\frac{1}{a} )$ and $\tilde M^{(8)}_{AS}$ is the mass of asymmetric kinks in the sectors $(-1,-\frac{1}{a})$ and $(\frac{1}{a},1)$. As Fig.~\ref{fig:modifiedphi8mass} shows,  at $a=3.243327$ all kinks of this model have the same mass, $\tilde M=0.117684$, and for values greater than this value of $a$, the mass of asymmetric kinks is greater than the mass of symmetric kinks. In fact, the parameter $a$ controls the mass ratio of symmetric and asymmetric kinks, which provides different cases in the scattering of kinks. In the limit of $a \to 1$, the asymmetric kink mass tends to zero. In the case of $a=1$, only the (symmetric) kink in the model has a mass of $\tilde M_S^{(8)}=\frac{2}{3} (3 \pi -8)$.

Apart from the mass, another quantity that is affected by the parameter $a$ is the stability potential. Considering small fluctuations around the static solution and substituting in the equation of motion, we arrive at the Schr\"odinger-like equation with the following effective potential

\begin{eqnarray}
\tilde U^{(8)}=\frac{d^2\tilde{V}^{(8)}(\phi)}{d\phi^2}\bigg{|}_{\phi=\tilde{\phi}^{(8)}(x)}.
\end{eqnarray}

Figs.~\ref{fig:modifiedphi8qmpS} and \ref{fig:modifiedphi8qmpAS} show the stability potential for different values of $a$ for symmetric and asymmetric kinks, respectively. The value of the potential at the limits of $x \to \pm \infty$ in both cases is $\frac{4(a-1)^2}{a^2}$. The Fig.~\ref{fig:modifiedphi8qmpS} depicts a volcano shaped Schr\"odinger-like potential with a deep minimum for small $a$ values. In particular, the potential with this behavior allows the formation of resonances, making it difficult for internal modes to be present. On the other hand, an increase in the parameter causes a change in the shape of the potential. We can observe an increase in the asymptotic maximum and a reduction in the potential minimum. Now, the linear perturbation analysis for the asymmetric kink is not symmetrical relative to the reflection $x\to -x$, see Fig.~\ref{fig:modifiedphi8qmpAS}. It is worth noting that increasing $a$ promotes the development of a maximum point in the potential as well as a rise in depth.

From translational invariance, we may infer the existence of the zero mode ($\omega_0=0$), which corresponds to the derivative of the solution, as follows $\eta_0(x)={d\phi}/{dx}$. We numerically investigated the occurrence of internal modes for both potentials and depict in Fig.~\ref{fig:modifiedphi8modes} the square frequency as a function of parameter $a$. We observe the occurrence of vibrational modes in asymmetric kink, as indicated by the red line. Moreover, the dashed black line represents the beginning of the continuous spectrum. It is worth noting that when $a$ increases, the frequency squared approaches the continuous mode. This behavior is associated with a change in the perturbation potential displayed in Fig.~\ref{fig:modifiedphi8qmpAS}. The increase in parameter $a$ shows the emergence of a thicker and deeper central well, favoring the appearance of bound states. However, for larger values of $a$, the Schr\"odinger-like potential no longer has equal asymptotic values, approaching a quantum mechanical potential that only presents the tranlational mode \cite{Dorey.PRL.2011}. In the symmetrical scenario, no bound state is observed for small values of $a$. In Fig. \ref{fig:modifiedphi8modes}, the blue line appears exclusively when $a>4.5$, indicating a single vibrational mode for the $\tilde{U}_S^{(8)}$. The absence of vibrational states for small values of this parameter is associated to the behavior of the perturbation potential. Notice that when $a$ decreases, the Schr\"odinger-like potential behaves as a volcano shaped potential. This type of potential favours the appearance of quasi-localized states. In addition to the analysis of the modes for an individual kink or antikink, we carried out the investigation for the collective kink-antikink and antikink-kink pairs for the asymmetric case. However, we discovered no additional states in any scenario.

In the following, we will discuss the kink-antikink and antikink-kink scattering process of the asymmetric and symmetric solution. Due to the non linearity of the model and solution behavior, the results are particularly complex. In addition, the modification of the parameter $a$ and the initial velocity produce intriguing effects. In this sense, we solved the equation of motion with $4^{th}$ order finite-difference method with a spatial step $\delta x=0.05$. For the time dependence we used a $6^{th}$ order symplectic integrator method with a time step $\delta t=0.02$. We fixed $x_0=\pm 10$ for the initial position of the pair.


\subsection{Asymmetric kink scattering}


In this section, we initially investigate the kink-antikink collision process of the solutions obtained from the deformation of the $\phi^4$ model. We first investigate the asymmetric kink collision $\tilde{\phi}^{(8)}_{AS}(\frac1a,1)$. For numerical solutions, we used the following initial conditions
\begin{eqnarray}
    \tilde\phi(x,x_0,v,0) &=& \tilde{\phi}_{AS}^{(8)}(x+x_0,v,0)-\tilde{\phi}_{AS}^{(8)}(x-x_0,-v,0)-1,\\
    \dot{\tilde \phi}(x,x_0,v,0) &=& \dot{\tilde{\phi}}_{AS}^{(8)}(x+x_0,v,0)-\dot{\tilde{\phi}}^{(8)}_{AS}(x-x_0,-v,0).
\end{eqnarray}
The time evolution in included with a boost for the static solutions, with $\gamma=(1-v^2)^{-1/2}$ and $v$ standing for the velocity. We mentioned before that large values of $a$ favor a potential $\tilde{U}^{(8)}_{AS}$ that is not symmetrical with respect to reflection. In this sense, the kink-antikink and the antikink-kink scattering are very different, so the initial conditions have a high sensitivity in the dynamics of the scattering.

\begin{figure*}[!ht]
\begin{center}
  \centering
\subfigure[ ]{\includegraphics[width=0.45\textwidth]{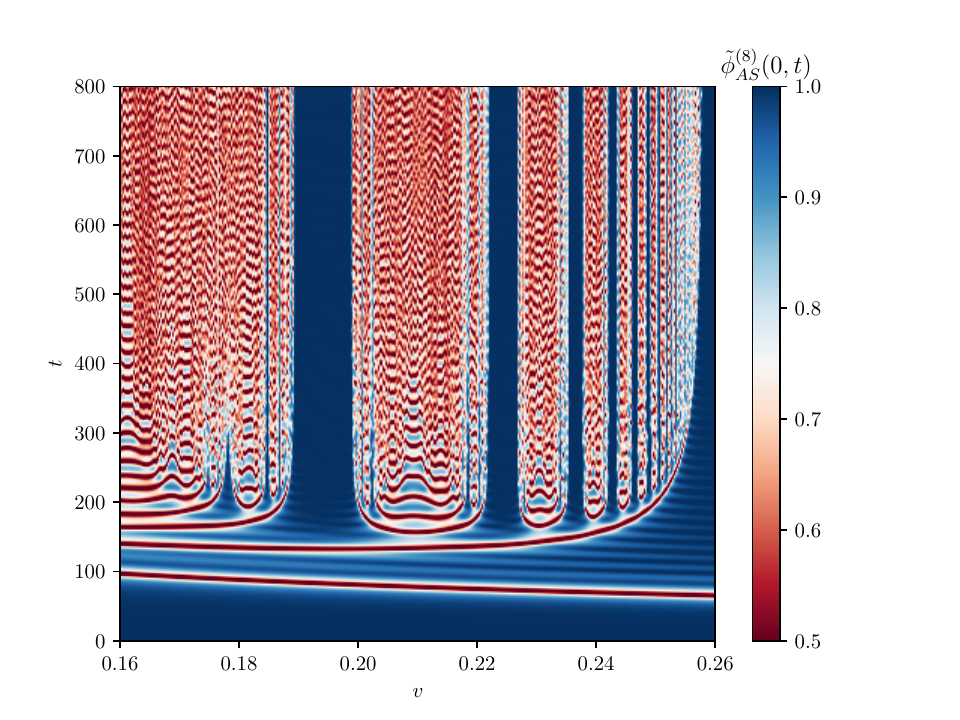}\label{s1a2times}}
\subfigure[ ]{\includegraphics[width=0.45\textwidth]{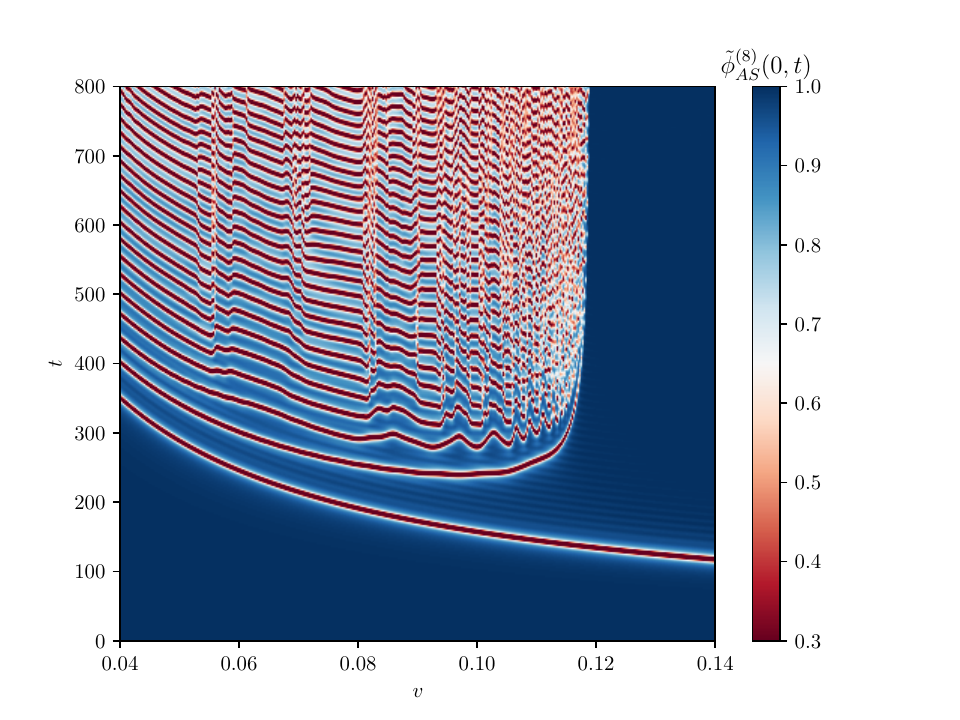}\label{s1a3times}}
  \caption{Kink-antikink - Evolution of the scalar field at the center of mass $\tilde{\phi}_{AS}^{(8)}(0,t)$ for the asymmetric kink located in the $(\frac{1}{a},1)$ sector as a function of time and initial velocity $v$ for (a) $a=2$ and (b) $a=3$. The colormap can be interpreted as follows: the red color represent the interaction between the kinks. In (a) the resonance windows (vertical blue regions) appear at the intervals when the time for the third collision diverges. In (b) the resonance windows are absent.}
\label{s1_timexv}
\end{center}
\end{figure*}

The structure of scattering for some values of $a$ is depicted in Fig. \ref{s1_timexv}. Figures \ref{s1a2times} and \ref{s1a3times} correspond to $a=2$ and $a=3$, respectively. It is worth noting that both have a single vibrational mode. For $a=2$, we observe the formation of two-bounce windows. This figure shows the evolution of scalar field at the center of mass as a function of time and initial velocity. The red lines correspond to the interactions between the kinks. Resonance windows are visible when we have only two red lines horizontally followed by a vertical blue region. When the second horizontal line diverges, the critical velocity is reached. For $a=3$, however, we do not notice the presence of the resonant windows. In this case, for $v<v_c\approx0.1203$, only bion states are achieved and for $v>v_c$, the output corresponds to inelastic scattering between the pair. 

\begin{figure*}[!ht]
\begin{center}
  \centering
\subfigure[ ]{\includegraphics[{angle=0,width=4cm,height=4cm}]{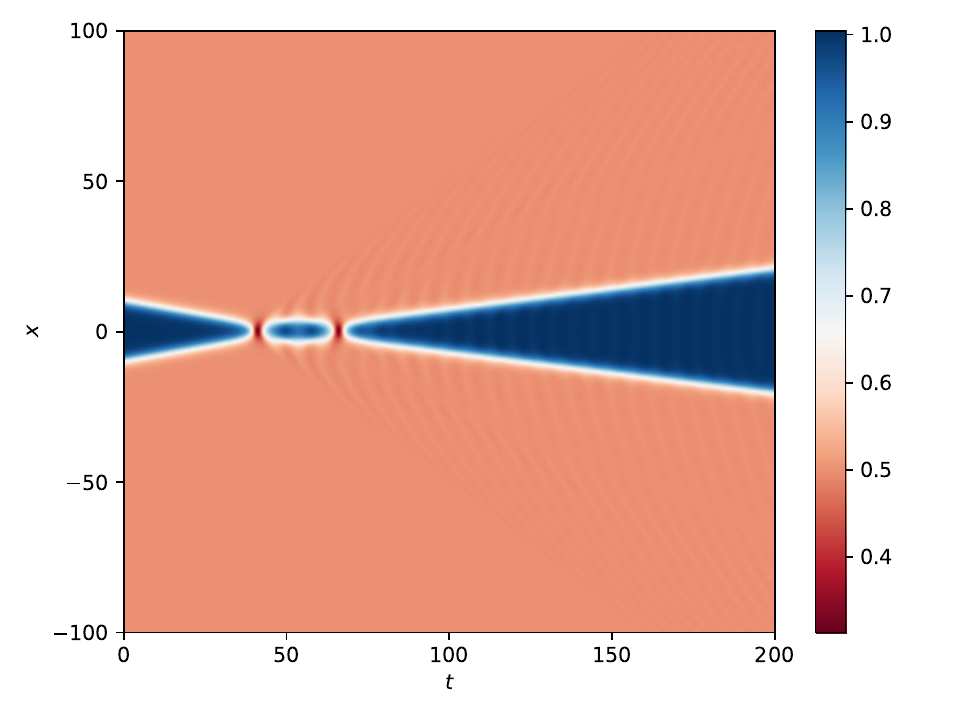}\label{s1a2A}}
\subfigure[ ]{\includegraphics[{angle=0,width=4cm,height=4cm}]{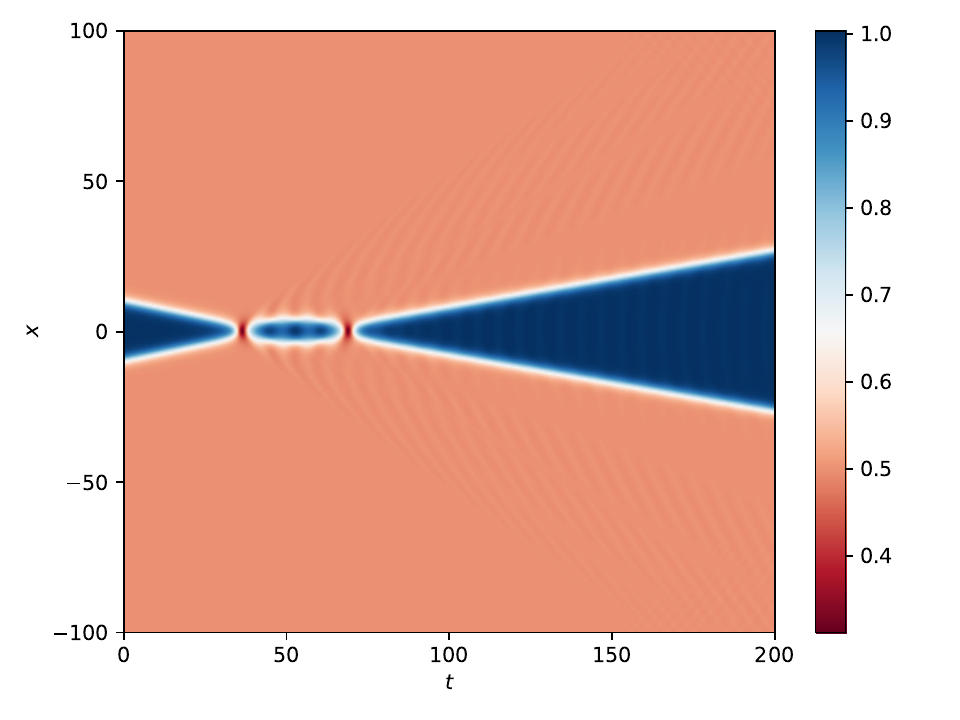}\label{s1a2B}}
\subfigure[ ]{\includegraphics[{angle=0,width=4cm,height=4cm}]{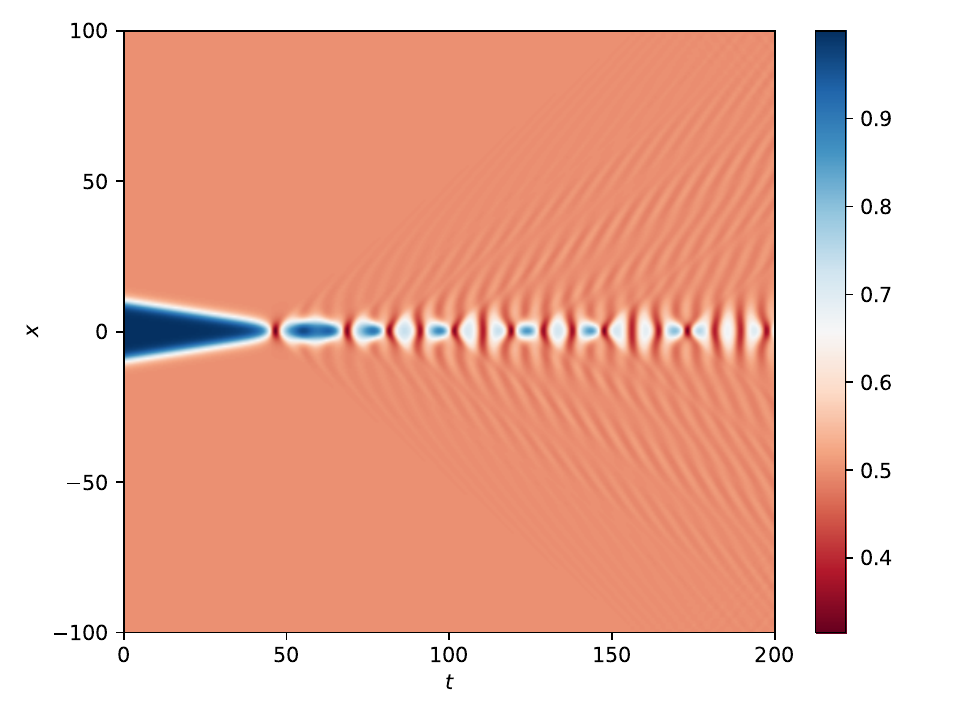}\label{s1a2C}}
\subfigure[ ]{\includegraphics[{angle=0,width=4cm,height=4cm}]{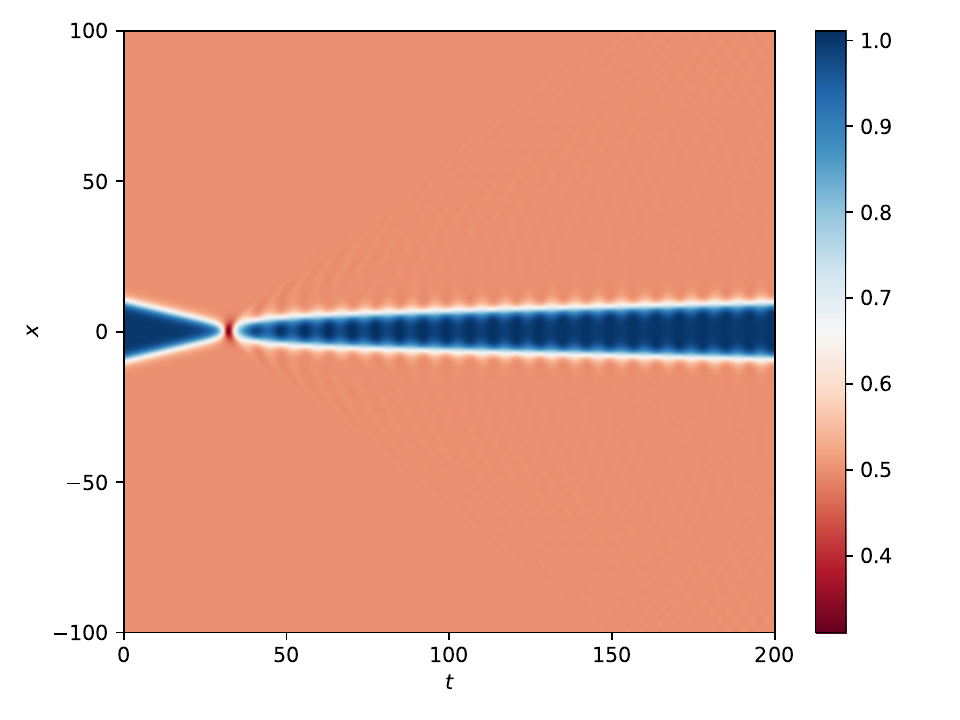}\label{s1a2D}}
  \caption{Kink-antikink - Evolution of scalar field $\tilde{\phi}^{(8)}_{AS}(\frac1a,1)$ in spacetime for $a=2$ with (a) $v=0.1930$, (b) $v=0.2240$, (c) $v=0.1660$ and (d) $v=0.2598$.}
\label{col_s1_a2}
\end{center}
\end{figure*}

In Fig. \ref{col_s1_a2}, some kink-antikink scattering results for the case where $a=2$ are shown. For instance, we depict the two-bounce scattering in Figs. \ref{s1a2A} and \ref{s1a2B}. In this case, the kink-antikink pair approaches and collides once, however, is unable to escape the mutual attraction, collides again a second time, and finally separates accompanied by a small radiation. The occurrence of these resonance windows is explained by the energy exchange mechanism between the zero and vibrational modes \cite{Campbell.PhysD.1983}. An important factor that distinguishes the location of each two-bounce window is the number $m$ of oscillations that occur between the bounces. For example, in Fig. \ref{s1a2A} for $v=0.1930$, we can see two oscillations between the collisions (the two blue dots), which corresponds to $m=2$, while in Fig.~\ref{s1a2B} for $v=0.2240$, we see $m=3$, which belongs to the third two-bounce window. Additionally, we observed the development of a state known as bion, see Fig.~\ref{s1a2C}, where the kink-antikink pair captures each other and becomes a trapped state. In this example, the scalar field at the center of mass undergoes considerable changes and oscillates erratically around the adjacent vacuum. Notably, the pair is annihilated after a prolonged time of scalar radiation emission. Finally, one-bounce scattering is depicted in Fig.~\ref{s1a2D}, where the kink approaches the antikink, collides only once, visits the adjacent vacuum and separates with a final velocity lower than the initial velocity.

\begin{figure*}[!ht]
\begin{center}
  \centering
\subfigure[ ]{\includegraphics[{angle=0,width=4cm,height=4cm}]{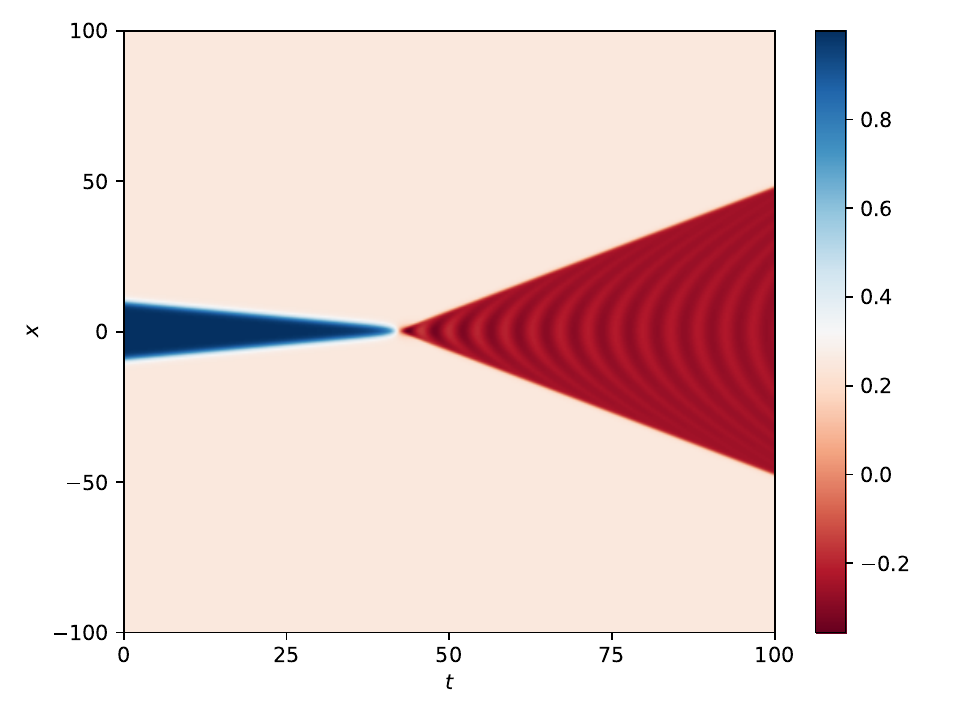}\label{s1a4}}
\subfigure[ ]{\includegraphics[{angle=0,width=4cm,height=4cm}]{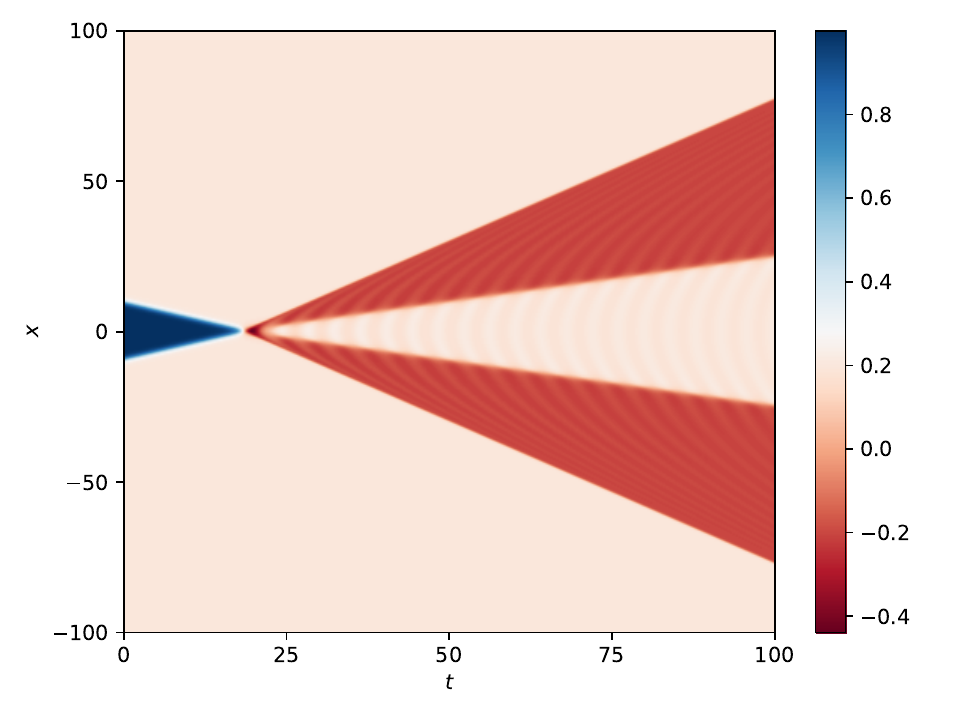}\label{s1a5}}
\subfigure[ ]{\includegraphics[{angle=0,width=4cm,height=4cm}]{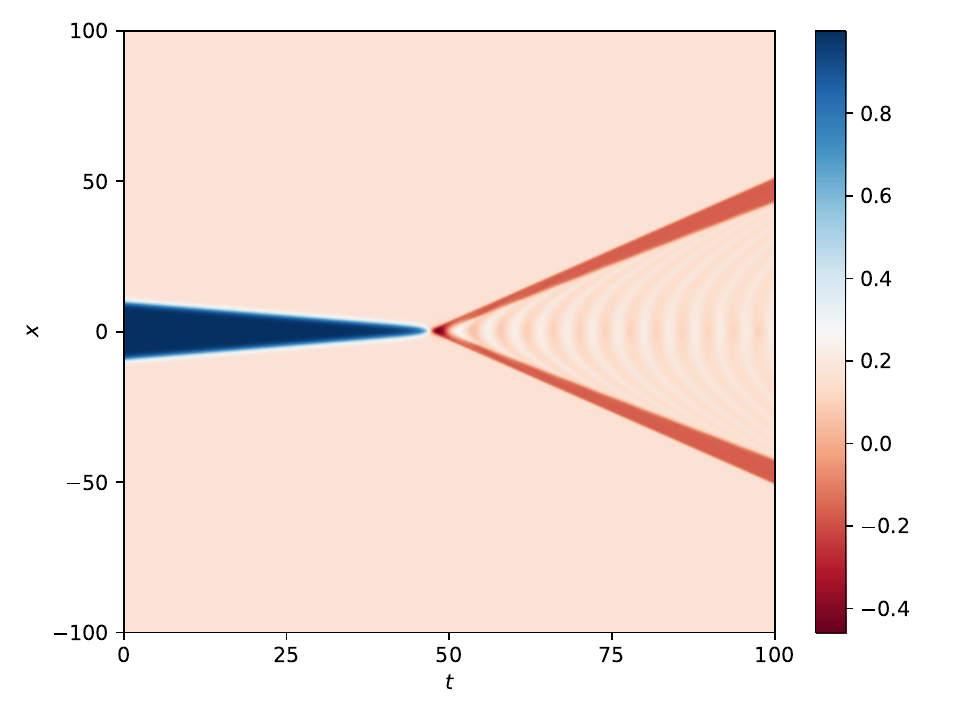}\label{s1a6}}
\subfigure[ ]{\includegraphics[{angle=0,width=4cm,height=4cm}]{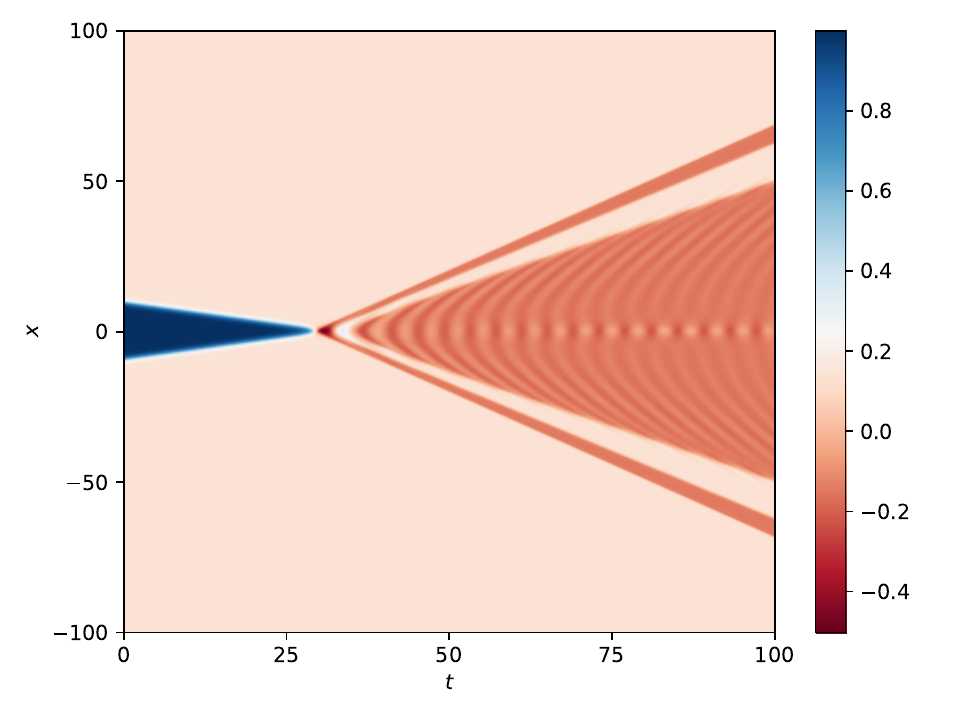}\label{s1a7}}
  \caption{Kink-antikink - Evolution of scalar field $\tilde{\phi}^{(8)}_{AS}(\frac1a,1)$ in spacetime for (a) $a=4$ with $v=0.20$, (b) $a=5$ with $v=0.50$, (c) $a=6$ with $v=0.18$ and (d) $a=7$ with $v=0.30$.}
\label{col_s1_a47}
\end{center}
\end{figure*}

An unusual pattern appears as the value of $a$ increases. The resonant windows are completely suppressed even when the internal mode is present, and the kink-antikink scattering results in an antikink-kink pair. In this sense, our numerical results for some values of $a$ are depicted in Fig. \ref{col_s1_a47}. We see that when $a=4$, the kink-antikink pair approaches, collides once and then separates on a new sector. We can realize the result of the asymmetric kink-antikink collision produces an antikink-kink pair that visits the symmetric sector. These behaviors are related to the massive character of the solutions. In the small $a$ region, the asymmetric kink is lighter than the symmetric kink, which makes it impossible to change sectors. However, as the parameter increases, the asymmetric kink becomes more massive, so sector switching becomes possible \cite{diojoaza}.

Notice that the increase in $a$ causes the resulting antikink-kink pair to form in a different topological sector - see Fig. \ref{s1a4}. Consequently, the parameter $a$ has a bigger influence on the development of new pairings, compared to the initial velocity. In particular, compare Fig. \ref{s1a5} and Fig. \ref{s1a6}. Although the latter has a small velocity, the outcome reveals the development of two pairs. In the case of Fig. \ref{col_s1_a47}, the asymmetric kink is heavier than the symmetric one, so the pair has enough energy to produce additional pairs in a sector different from the initial one. The generation of new pairs was also seen in Ref. \cite{simas.2020}, where large kink scattering produces new pairs. The presence of simply the zero mode, followed by the continuum mode, is reported by the authors. On the other side, the model considered in this paper, has a zero mode and an internal mode for the asymmetric kink. Additionally, we further observe the appearance of an oscillating pulse after the collision at $x=0$ - see Fig. \ref{s1a7}. We discovered in our numerical simulations that a higher initial velocity promotes the emergence of an oscillating pulse at the collision center, consequently, a new pair must be produced from this perturbation. This phenomenon of oscillations at the collision center causing the development of new kink pairs was also described in Ref. \cite{roman.2010}.

In the following, we shall now consider antikink-kink scattering, due of the asymmetric nature of the solution $\tilde{\phi}^{(8)}_{AS}(\frac1a,1)$. We used the following initial conditions
\begin{eqnarray}
    \tilde\phi(x,x_0,v,0) &=& \tilde{\phi}_{AS}^{(8)}(x-x_0,-v,0)-\tilde{\phi}_{AS}^{(8)}(x+x_0,v,0)-\frac{1}{a},\\
    \dot{\tilde\phi}(x,x_0,v,0) &=& \dot{\tilde{\phi}}_{AS}^{(8)}(x-x_0,-v,0)-\dot{\tilde{\phi}}^{(8)}_{AS}(x+x_0,v,0),
\end{eqnarray}
where $\tilde\phi(x,x_0,v,t)=\tilde\phi(\gamma(x-vt))$ means a boost for the static solution with $\gamma=(1-v^2)^{-1/2}$. 

\begin{figure*}[!ht]
\begin{center}
  \centering
\subfigure[ ]{\includegraphics[{angle=0,width=7cm,height=5cm}]{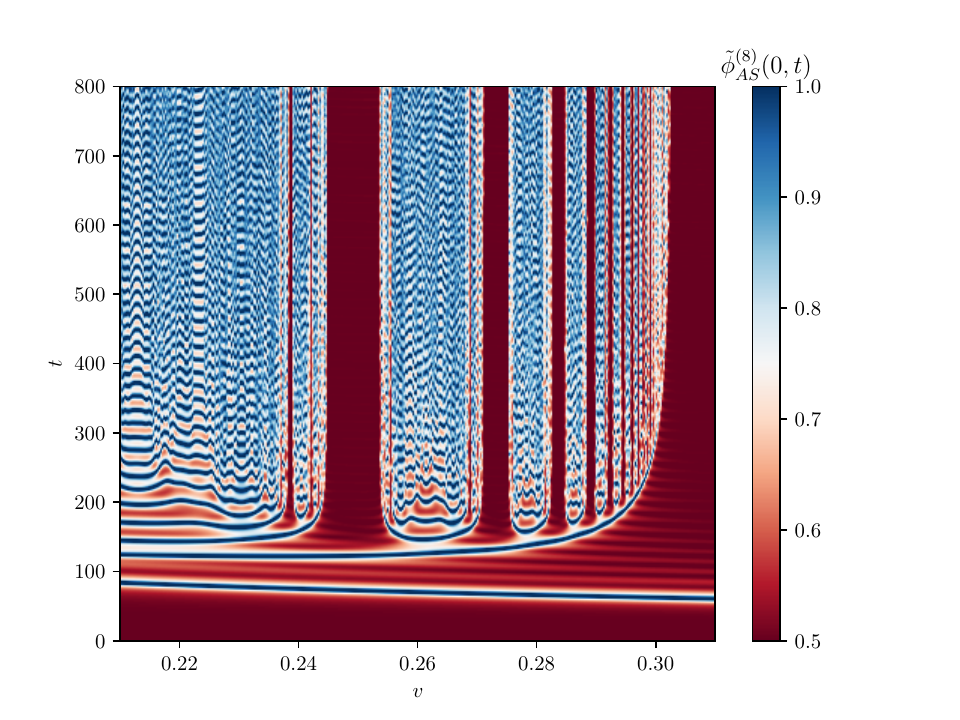}\label{s1a2timesAK}}
\subfigure[ ]{\includegraphics[{angle=0,width=7cm,height=5cm}]{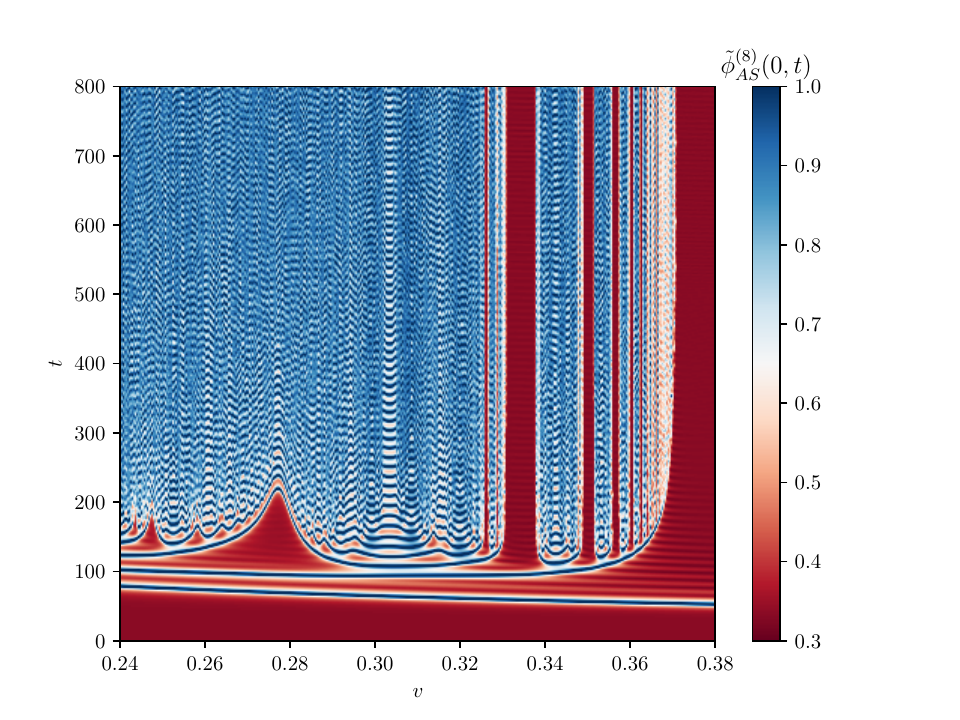}\label{s1a3timesAK}}
\subfigure[ ]{\includegraphics[{angle=0,width=7cm,height=5cm}]{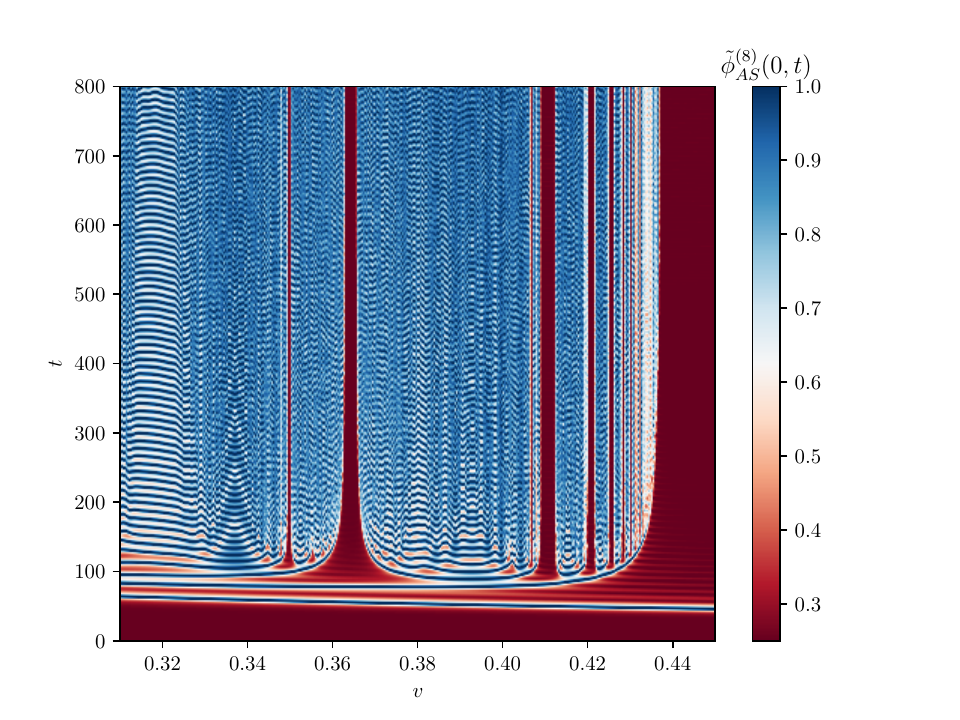}\label{s1a4timesAK}}
\subfigure[ ]{\includegraphics[{angle=0,width=7cm,height=5cm}]{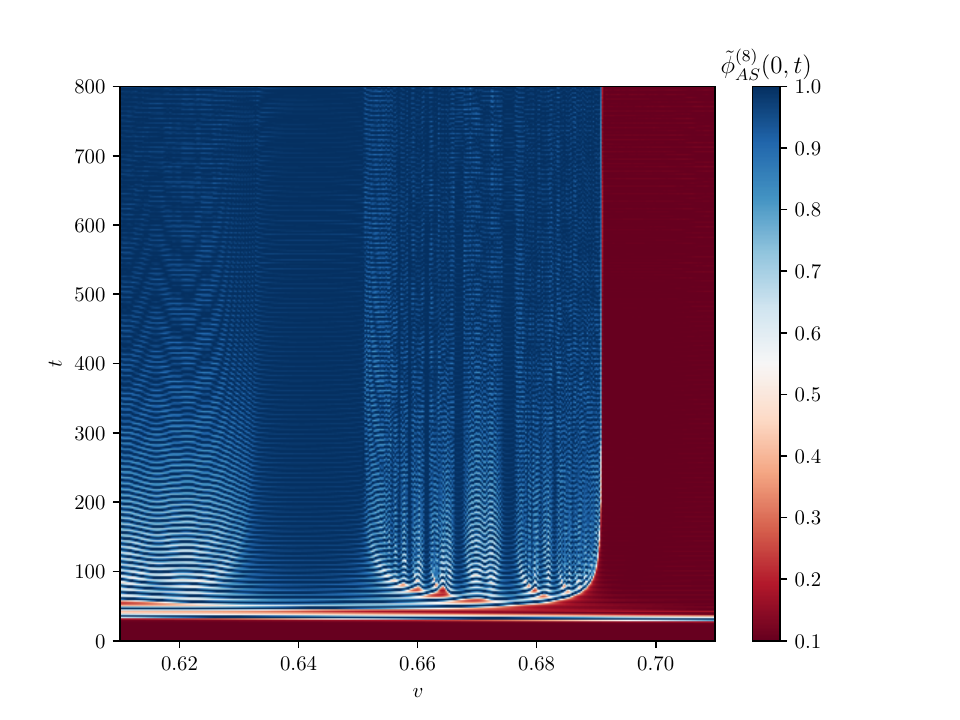}\label{s1a10timesAK}}
  \caption{Antikink-kink - Evolution of the scalar field at the center of mass $\tilde{\phi}_{AS}^{(8)}(0,t)$ for the asymmetric kink located in the $(\frac{1}{a},1)$ sector as a function of time and initial velocity $v$ for (a) $a=2$, (b) $a=3$, (c) $a=4$ and (d) $a=10$. The colormap can be interpreted as follows: the blue lines represent the interaction between the kinks.  In (a), (b) and (c) the resonance windows (vertical red region) appear at the intervals when the time for the third collision diverges. In (d) the resonance windows are absent.}
\label{s1_timexvAK}
\end{center}
\end{figure*}

We present a summary of the antikink-kink interaction findings for certain values of $a$ in Fig. \ref{s1_timexvAK}. In this figure we show the evolution of the scalar field as a function of initial velocity and time. The collisions are illustrated by the horizontal blue lines and the two-bounce windows are distinguished by their diverging bands. For small values of $a$, we can observe the presence of resonant windows. However, as the value of this parameter increases, the number of windows decreases until they are completely suppressed - see Fig. \ref{s1a2timesAK}-\ref{s1a10timesAK}. Additionally, the disappearance of the first two-bounce windows leads to the emergence of false resonance windows. The peak at $v\approx0.28$ in Fig. \ref{s1a3timesAK} represents this false window behavior. In this case, the antikink-kink pair approaches, collides twice and separates. However, after one period, it collides again and forms an oscillatory state. The appearance of false two-bounce windows was reported in Ref. \cite{Dorey.PRL.2011}. Furthermore, increasing this parameter contributes to an increase in the critical velocity, which leads to a larger region with bion states.

\begin{figure*}[!ht]
\begin{center}
  \centering
\subfigure[ ]{\includegraphics[{angle=0,width=4cm,height=3cm}]{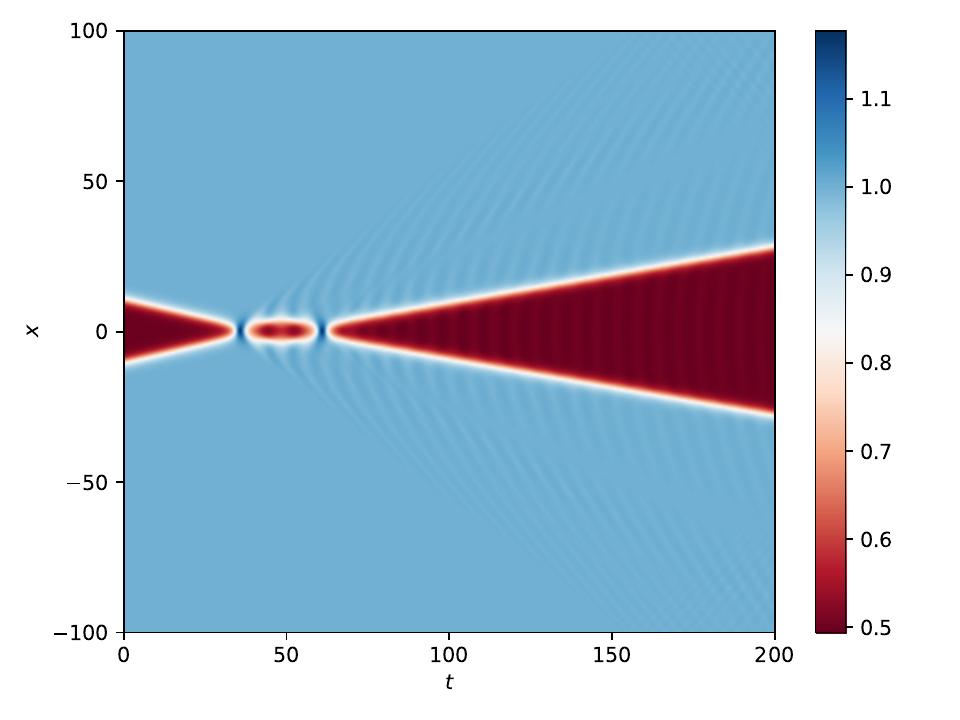}\label{ak_s1a2}}
\subfigure[ ]{\includegraphics[{angle=0,width=4cm,height=3cm}]{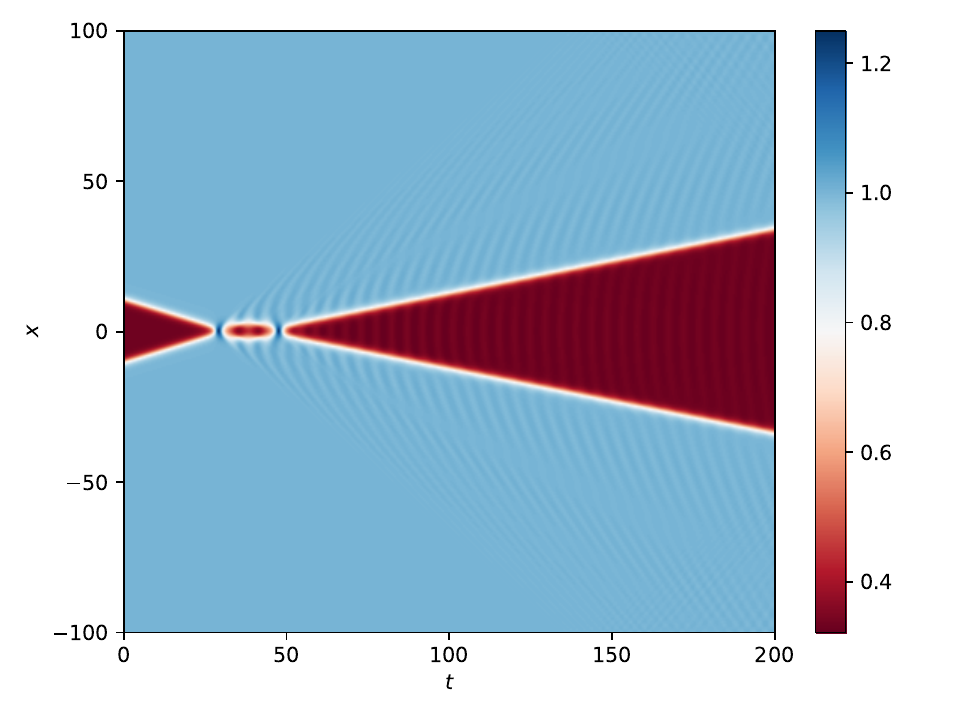}\label{ak_s1a3}}
\subfigure[ ]
{\includegraphics[{angle=0,width=4cm,height=3cm}]{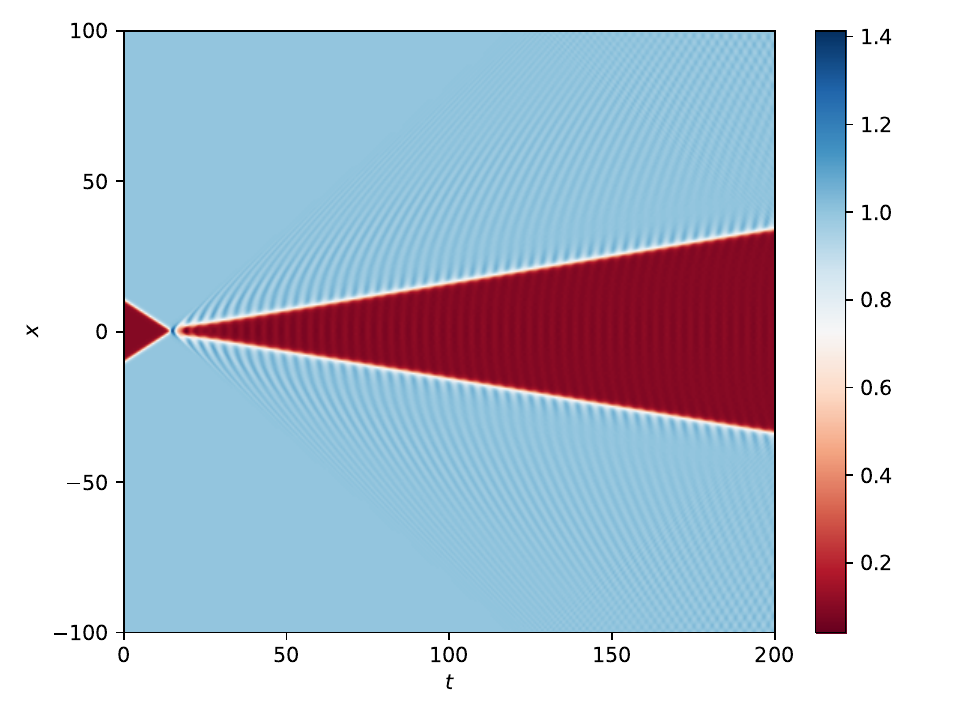}\label{ak_s1a10}}
\subfigure[ ]
{\includegraphics[{angle=0,width=4cm,height=3cm}]{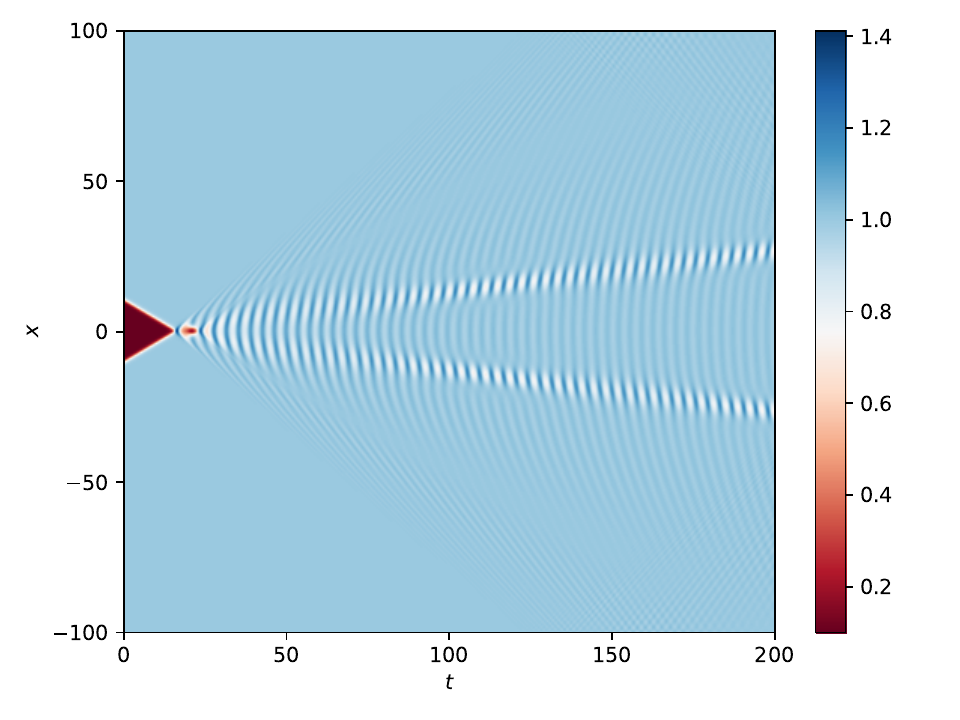}\label{ak_s1a102}}
  \caption{Antikink-kink - Evolution of scalar field $\tilde{\phi}^{(8)}_{AS}(\frac1a,1)$ in spacetime for (a) $a=2$ with $v=0.25$, (b) $a=3$ with $v=0.3320$ and $a=10$ with (c) $v=0.70$ and (d) $v=0.64$.}
\label{col_ak_s1_a2_10}
\end{center}
\end{figure*}

Unlike the asymmetric kink-antikink collision, the antikink-kink configuration does not allow the pair to visit the symmetric vacuum, as we can seen in the Fig. \ref{ak_s1a2} and Fig. \ref{ak_s1a3}. We show the occurrence of two-bounce for $a=2$ and $a=3$, respectively and the one-bounce behavior for $a=10$ (Fig. \ref{ak_s1a10}). Furthermore, we report the production of oscillating pulses for $a=10$ - see Fig. \ref{ak_s1a102}. Similar solutions were discovered in Ref. \cite{roman.2010}. As a result of the collision of the antikink-kink pair, we observe two waves that scatter almost harmonically.

The reason why kink-antikink and antikink-kink scattering yield different results has to do with how the solutions are placed on the line that defines the initial configuration. For instance, in the kink-antikink collision, the region connecting the kink to the antikink does not vary with the change in $a$. However, in the antikink-kink case, changes in $a$ alter the tails of the antikink and kink. Moreover, the suppression of the resonance windows for the collision of the asymmetric solution can be explained by the presence of the set of internal modes of the model. For small values of $a$, the asymmetric kink has just one internal mode. However, when the value increases, it favors the occurrence of a vibrational state for the symmetric kink. As a result, the internal mode of the symmetric solution influences the vibrational state of the asymmetric kink. Consequently, the mechanism for exchanging resonant energy between translational and vibrational modes become frustrated.

\subsection{Symmetric kink scattering}


Now, we investigate the kink-antikink collision process of the symmetric solutions $\tilde{\phi}^{(8)}_{S}(-\frac1a,\frac1a)$. The kink-antikink and antikink-kink scattering both generate the same outcomes. For this reason, we shall exclusively study the kink-antikink collision. We used the following initial conditions
\begin{eqnarray}
    \tilde\phi(x,x_0,v,0) &=& \tilde{\phi}_{S}^{(8)}(x+x_0,v,0)-\tilde{\phi}_{S}^{(8)}(x-x_0,-v,0)-\frac1a,\\
    \dot{\tilde \phi}(x,x_0,v,0) &=& \dot{\tilde{\phi}}_{S}^{(8)}(x+x_0,v,0)-\dot{\tilde{\phi}}^{(8)}_{S}(x-x_0,-v,0),
\end{eqnarray}
where $\tilde\phi(x,x_0,v,t)=\tilde\phi(\gamma(x-vt))$ means a boost for the static solution with $\gamma=(1-v^2)^{-1/2}$. In contrast to the preceding case, the symmetrical kink consistently possesses a symmetrical potential $\tilde{U}^{(8)}_S$ concerning the reflections. Thus, regardless of the order in which the initial condition is placed, the results do not differ.

\begin{figure*}[!ht]
\begin{center}
  \centering
\subfigure[ ]{\includegraphics[{angle=0,width=4cm,height=4cm}]{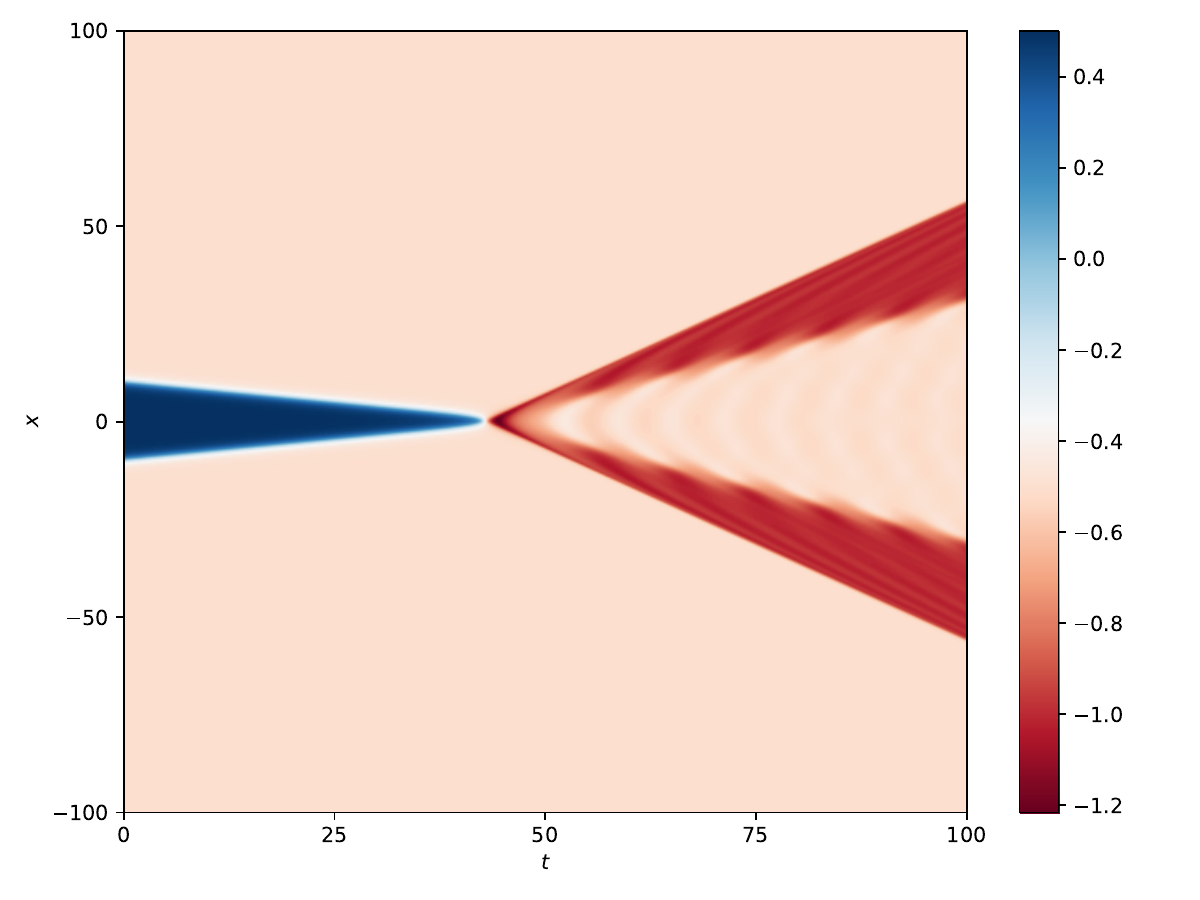}\label{ka_s2a2}}
\subfigure[ ]{\includegraphics[{angle=0,width=4cm,height=4cm}]{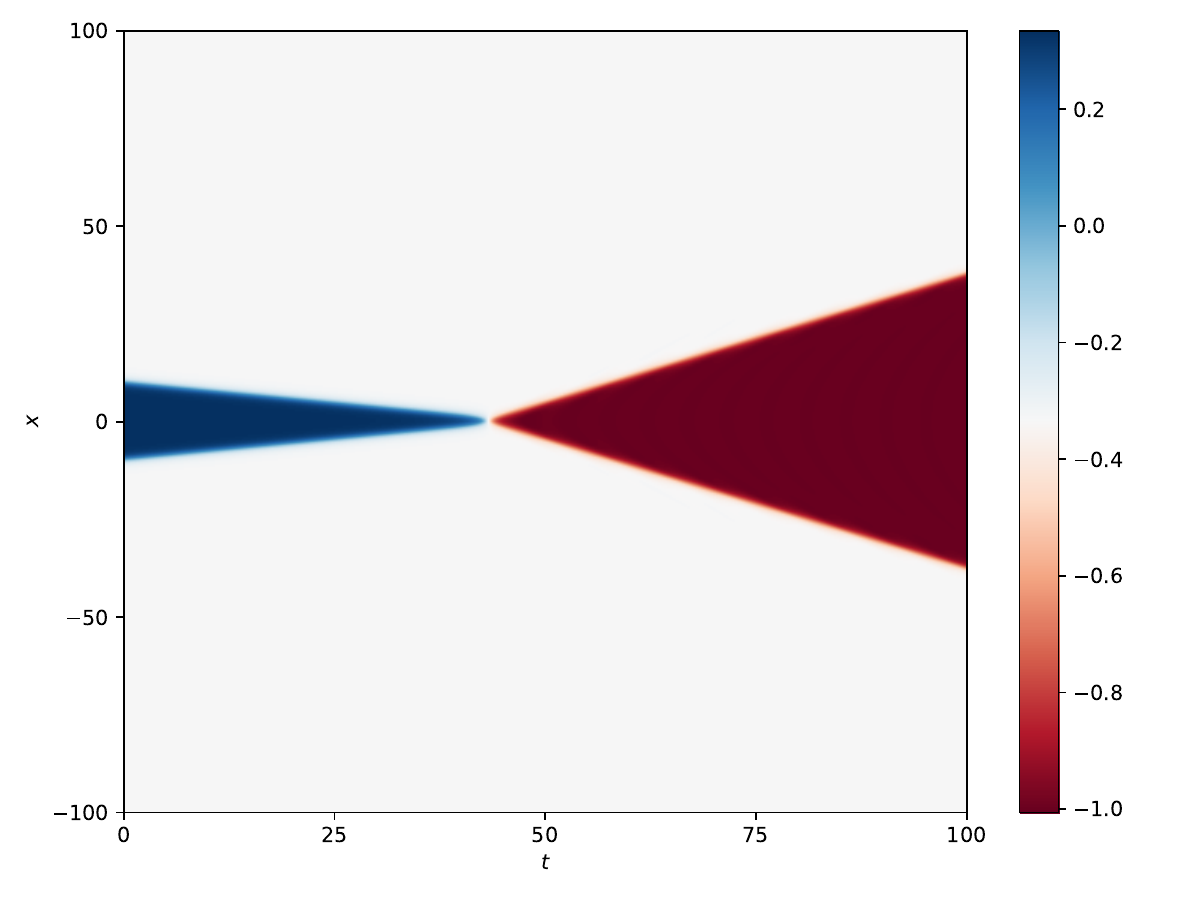}\label{ka_s2a3}}
\subfigure[ ]
{\includegraphics[{angle=0,width=4cm,height=4cm}]{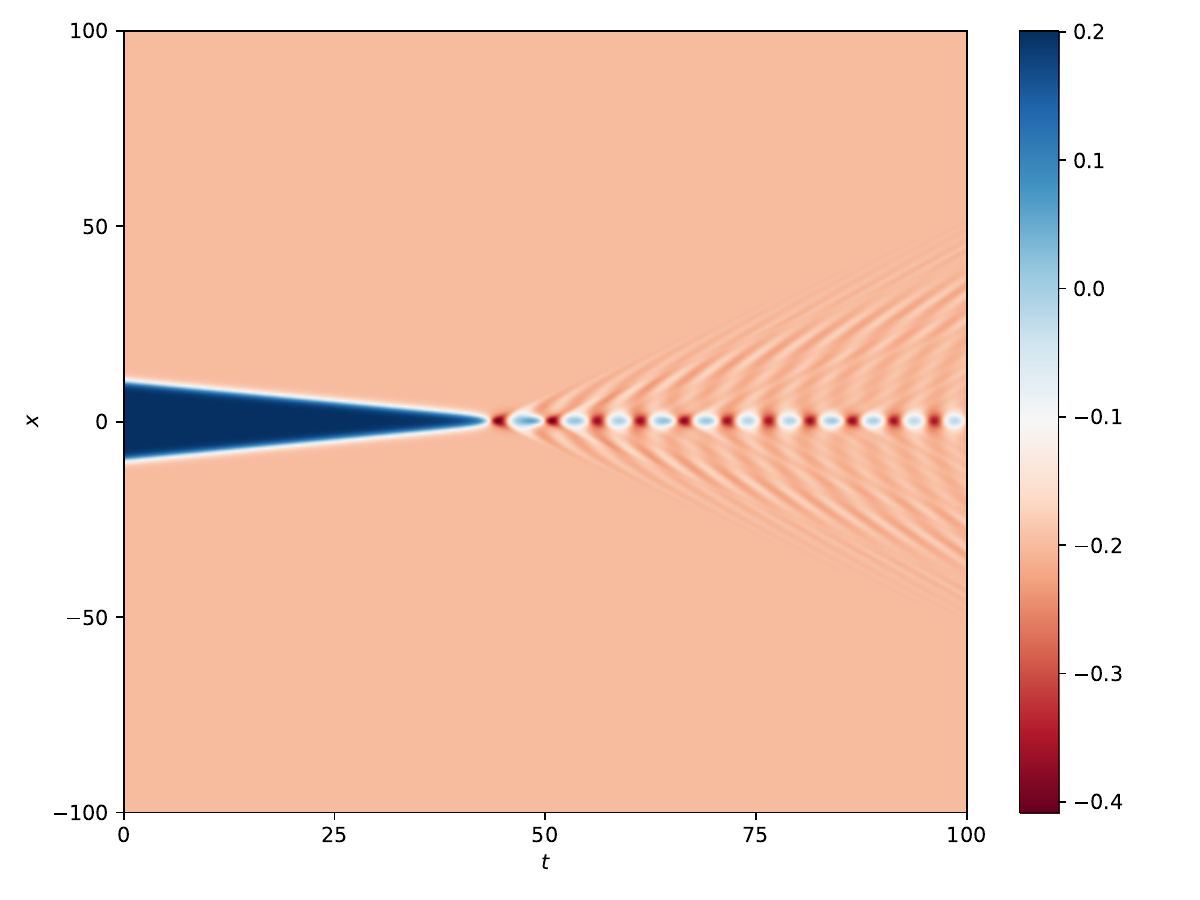}\label{ka_s2a5}}
  \caption{Kink-antikink - Evolution of scalar field $\tilde{\phi}^{(8)}_{S}(-\frac1a,\frac1a)$ in spacetime for (a) $a=2$, (b) $a=3$ and (c) $a=5$ with $v=0.20$.}
\label{col_ka_s2_a}
\end{center}
\end{figure*}

In the first moment, we examine the scattering by altering the parameter $a$ while keeping the initial velocity constant. Some of these examples are illustrated in Fig. \ref{col_ka_s2_a}. For $a=2$, we see the production of two antikink-kink pairs after collision. Importantly, the scattered pairs are asymmetric kinks in the topological sector $(-\frac1a,-1)$. In addition, the increase in $a$ results in the development of a single antikink-kink pair, as reported in Fig. \ref{ka_s2a3} for $a=3$. It is worth noting that in this region, the symmetrical kink is more massive and becomes lighter as $a$ increases. There are no two-bounce windows or bion development in this region of $a$ values. On the other hand, larger values of $a$ modify this pattern and reveal the presence of a bion at the center of the collision, as seen in Fig. \ref{ka_s2a5} for $a=5$. The emergence of the vibrational state for symmetric kink is associated with the alteration in behavior.

\begin{figure*}[!ht]
\begin{center}
  \centering
\subfigure[ ]{\includegraphics[{angle=0,width=4cm,height=4cm}]{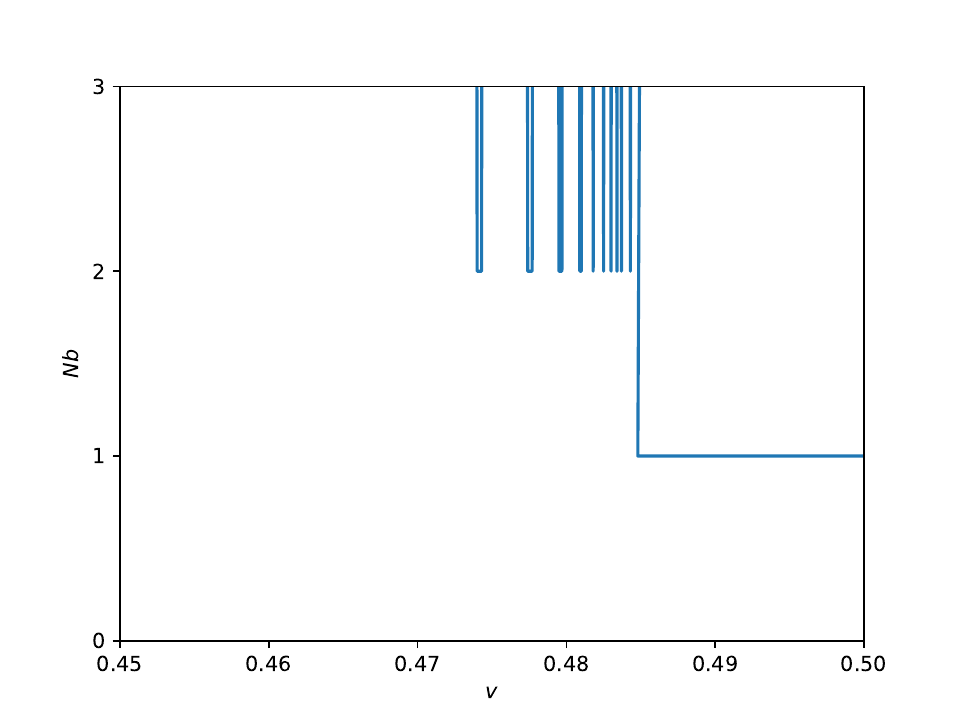}\label{ka_Nbxv_s2a6}}
\subfigure[ ]{\includegraphics[{angle=0,width=4cm,height=4cm}]{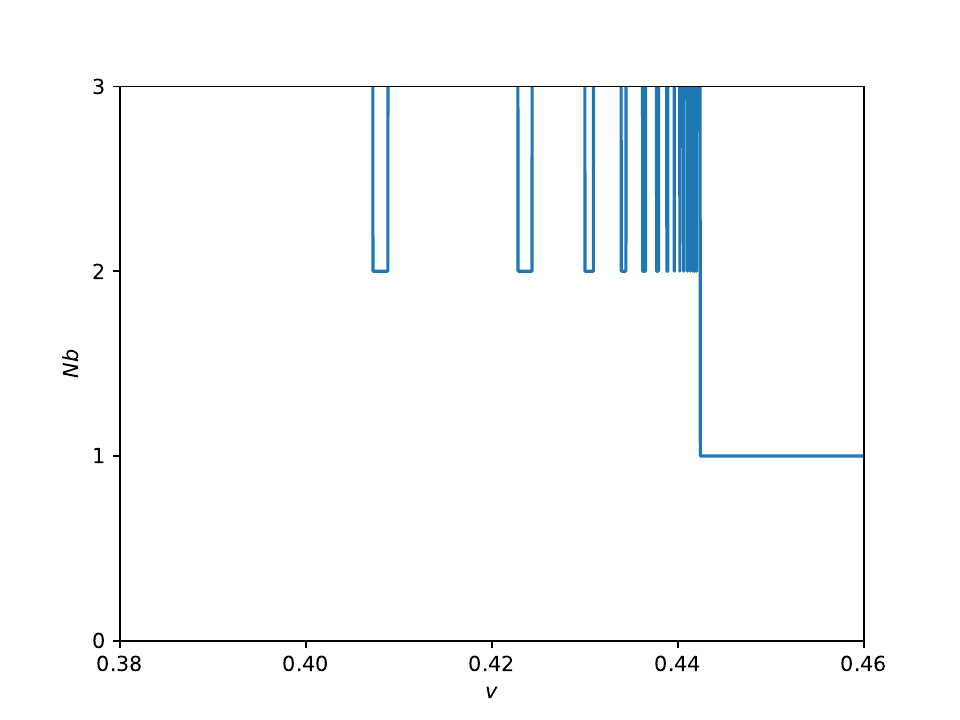}\label{ka_Nbxv_s2a7}}
\subfigure[ ]
{\includegraphics[{angle=0,width=4cm,height=4cm}]{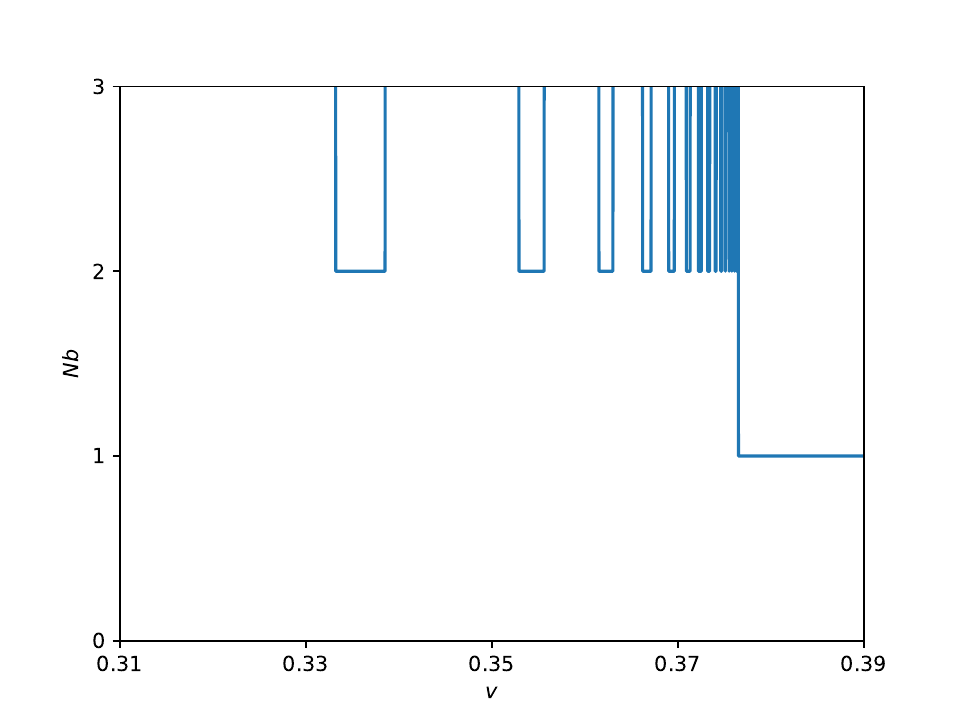}\label{ka_Nbxv_s2a10}}
{\includegraphics[{angle=0,width=4cm,height=4cm}]{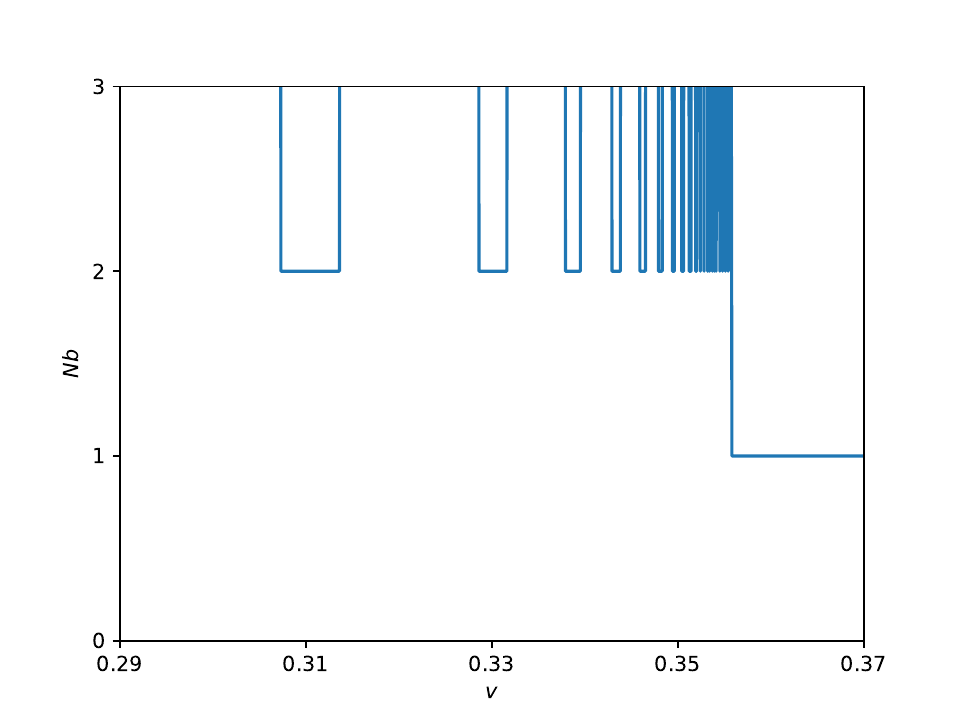}\label{ka_Nbxv_s2a12}}
  \caption{Kink-antikink - Number of bounces versus initial velocity of scalar field $\tilde{\phi}^{(8)}_{S}(-\frac1a,\frac1a)$ for (a) $a=6$, (b) $a=7$, (c) $a=10$ and (d) $a=12$.}
\label{ka_Nbxv_s2_a}
\end{center}
\end{figure*}

In order to better understand the impact of the parameter $a$, we developed a outline for identifying the formation of the two-bounce windows by looking at the number of bounces as a function of the initial velocity. This resonant structure, as we know, may be constructed by counting the bounces of the scalar field at the center of mass $\phi(0,t)$. We only include the structure corresponding to one and two collisions in Fig. \ref{ka_Nbxv_s2_a}. The absence of any regions suggests either the production bion or higher order collisions. In regions with high velocities, only inelastic collisions are observed $(Nb=1)$. Furthermore, for intermediate velocities, peaks can be seen in the pictures depicting the region corresponding to two collisions $(Nb=2)$. In our investigation, we see that when the parameter increases, the number of two-bounce windows increases. Specifically, with $a=5$, we observe the formation of only a few windows. However, for $a=12$, the number of windows is higher, and these become wider. Clearly, the model contains two vibrational modes for both symmetric and asymmetric kink at values around $a\sim 4.6$. The presence of this set of internal modes in the model difficulties the energy exchange mechanism during the collision. Consequently, creating resonant windows is more challenging. On the other hand, increasing $a$ shows that $\omega^2_1$, for the asymmetric solution, tends to approach the continuous mode even more. Thus, for larger values of the parameter, there is a greater contribution from the vibrational mode of the symmetric kink. Therefore, the energy exchange mechanism between the translational and vibrational modes is more effective, and there is an increase in the number of two-bounce windows. In addition, with $a\rightarrow \infty$, the deformed $\phi^8$ tends towards the $\phi^4$ model. One effect of this outcome is decrease in critical velocity. The critical velocity approaches $v_c \approx 0.2598$ as $a$ increases.

\section{modified \texorpdfstring{$\phi^{10}$}{pdfbookmark} model}
\label{sec:modifiedphi10model}

\begin{figure*}[!ht]
\begin{center}
  \centering
    \subfigure[\quad Deformation function]{\includegraphics[width=0.45
 \textwidth]{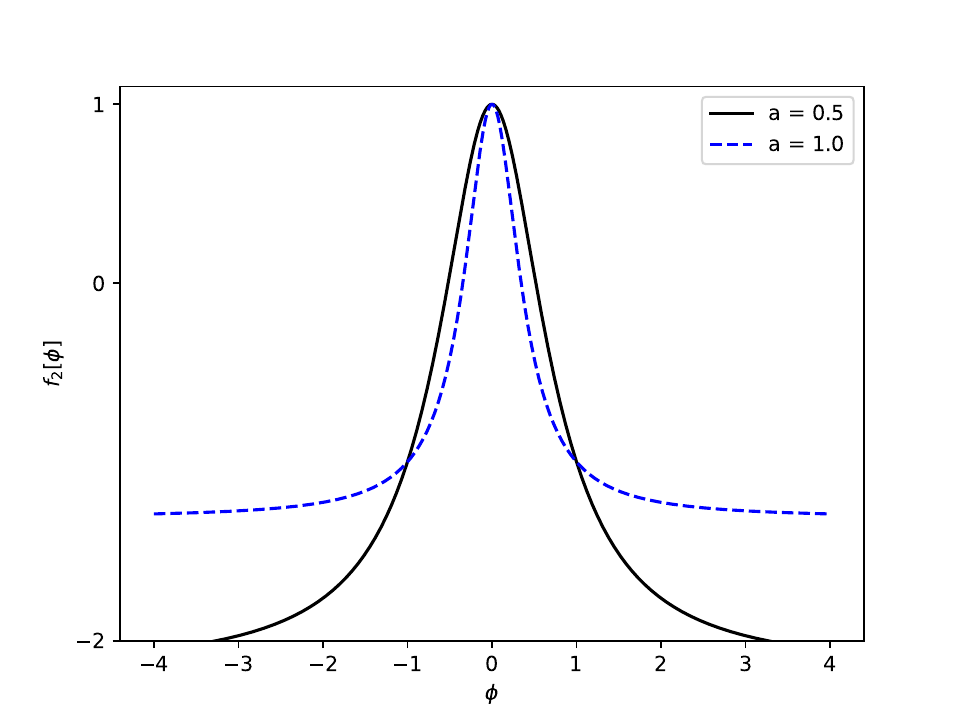}\label{fig:f2}}
  \subfigure[\quad Potentials]{\includegraphics[width=0.45
 \textwidth]{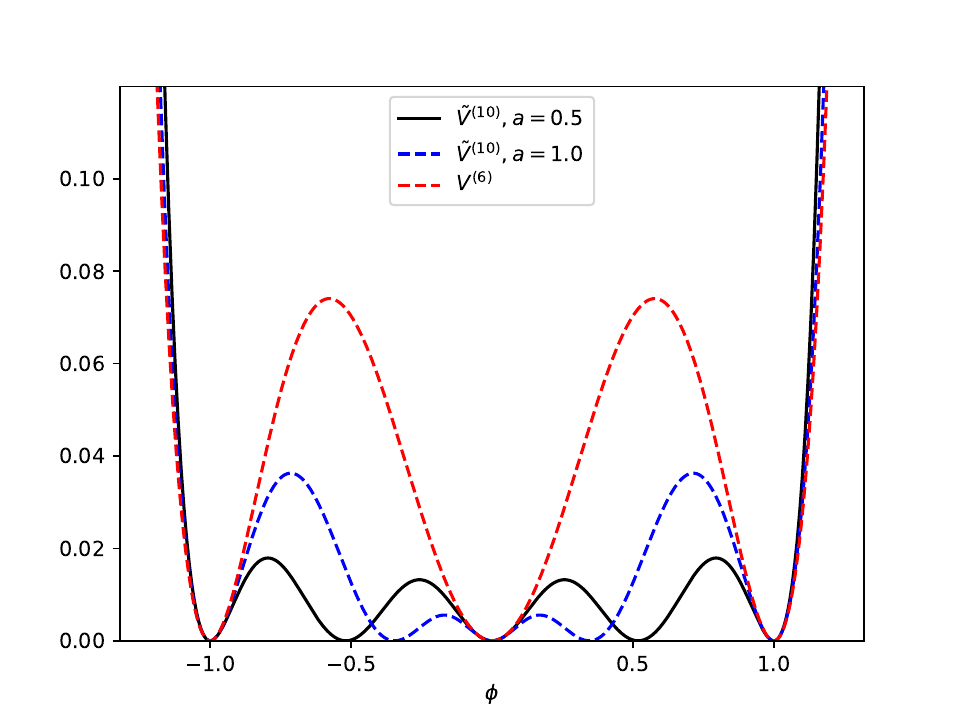}\label{fig:modifiedphi10potentials}}
\\
  \subfigure[\quad Modified $\phi^{10}$ kinks for $a=0.5$]{\includegraphics[width=0.45
 \textwidth]{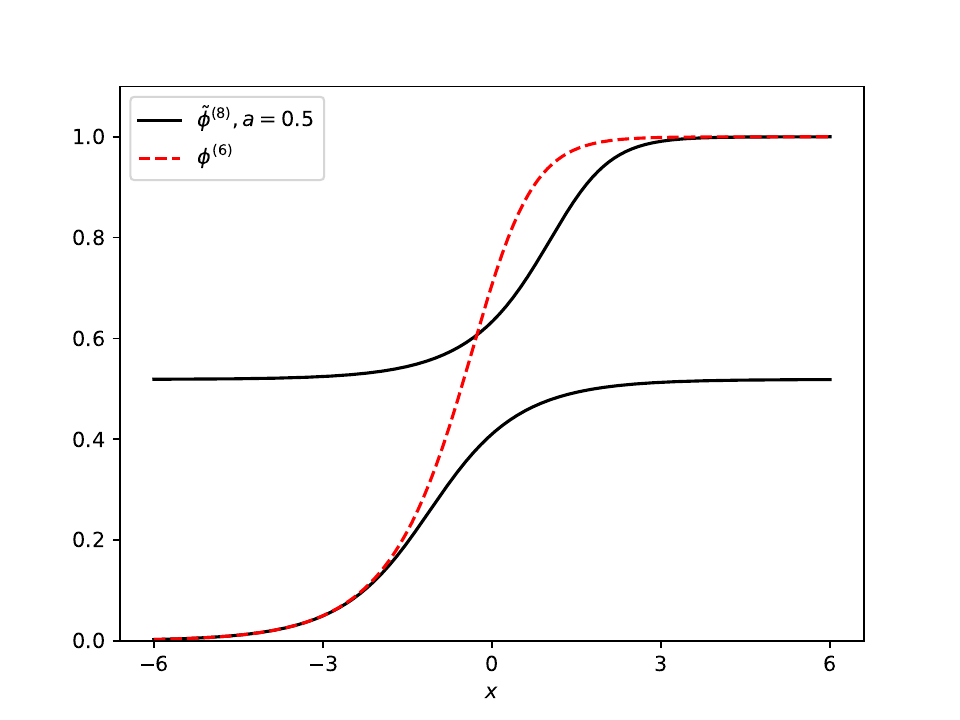}\label{fig:modifiedphi10kinksaLnSqrt3}}
  \subfigure[\quad Modified $\phi^{10}$ kinks for $a=1.0$]{\includegraphics[width=0.45
 \textwidth]{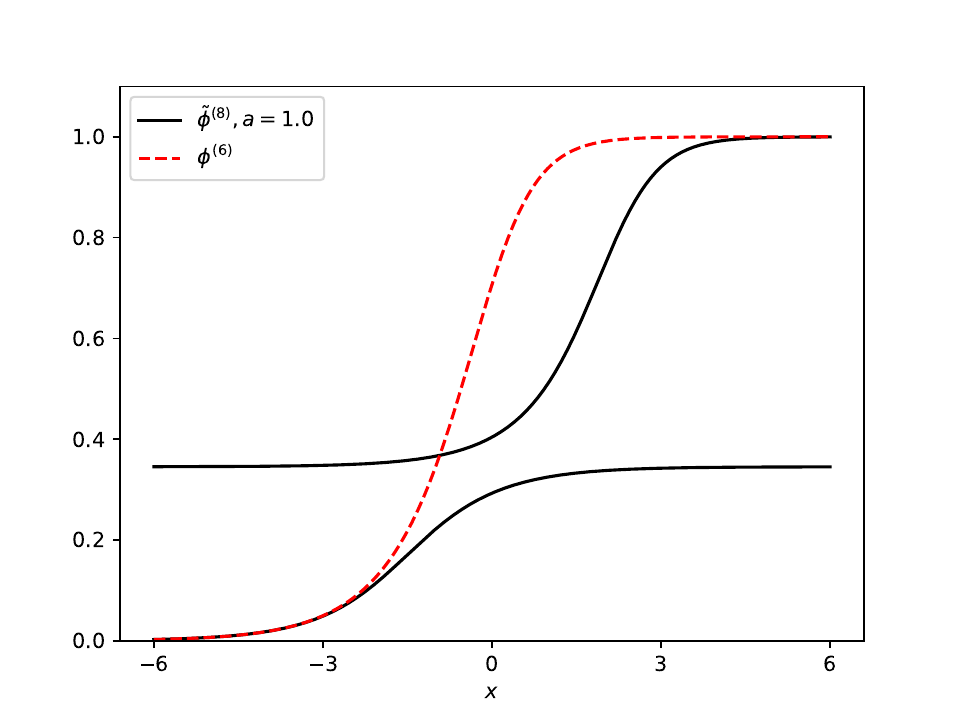}\label{fig:modifiedphi10kinksaLnSqrt8}}
\\
  \caption{(a) The deformation function $f_2[\phi]$ as a function of $\phi$, (b) modified $\phi^{10}$ potential and modified $\phi^{10}$ kinks for (c) $a=0.5$ and (d) $a=1.0$. The $\phi^6$ model is represented by the dashed red line.}
\end{center}
\end{figure*}

Second, we introduce the deformation function $f_2[\phi]$, as we can see in Fig.~\ref{fig:f2}

\begin{equation}\label{eq:f2}
f_2[\phi]=\frac{1-(1+e^{2 a} )\phi ^2}{1-(1-e^{2 a} )\phi ^2}.
\end{equation}
Here, $a$ is a real parameter. By using of Eq.~\eqref{eq:deformedpotential}, this deformation function can change the $\phi^4$ model with the potential $V^{(4)}=\frac{1}{2}(1-\phi^2)^2$ to the $\phi^6$ model with the potential $V^{(6)}=\frac{1}{2}\phi^2(1-\phi^2)^2$. We can use again $f_2[\phi]$ in $\phi^6$ model and find the following modified $\phi^{10}$ model,

\begin{eqnarray}\label{eq:modifiedphi10potential}
\tilde{V}^{(10)}&=& \frac{\phi ^2 \left(\phi ^2-1\right)^2 \left(\left(e^{2 a}+1\right) \phi ^2-1\right)^2}{2 \left(\left(e^{2 a}-1\right) \phi ^2+1\right)^2}.
\end{eqnarray}

This potential has five minima, at $0,\pm \frac{1}{\sqrt{1+e^{2a}}},$ and $\pm 1$. The behavior of the potential is shown in Fig.~\ref{fig:modifiedphi10potentials}. We consider positive values for the parameter $a$. 

For this model, the kink solutions are written as

\begin{eqnarray}\label{eq:modifiedphi10kinks}
\tilde{\phi}^{(10)} &=&
\begin{cases}
\tilde{\phi}_{(0,\frac{1}{\sqrt{e^{2a}+1}})}^{(10)}=\sqrt{\frac{2e^{2 x}}{2 \left(e^{2 a}+1\right) e^{2 x}+\sqrt{4 e^{2 (a+x)}+1}+1}}, \\
\tilde{\phi}_{(\frac{1}{\sqrt{e^{2a}+1}},1)}^{(10)}=\frac{1}{\sqrt{2}}\sqrt{\frac{\sqrt{4 e^{2 (a-x)}+1}+2 e^{2 (a-x)}+2 e^{-2 x}+1}{2 e^{2 (a-x)}+e^{4 a-2 x}+e^{-2 x}+1}}.
\end{cases}
\end{eqnarray}
In Figs. \ref{fig:modifiedphi10kinksaLnSqrt3} and \ref{fig:modifiedphi10kinksaLnSqrt8} one shows these kinks for two different values of parameter $a$. The configuration of the solution reveals a kink that varies between the minima $\frac{1}{\sqrt{e^{2a}+1}}$ and $1$, which we refer to as up kink. In addition to this, we can also see another kink between $0$ and $\frac{1}{\sqrt{e^{2a}+1}}$, which will be referred to here as down kink. The behavior of collision will also depend on the initial configuration, therefore, kink-antikink scattering differs from antikink-kink collision. 

In particular, at $a \to \infty$, the modified $\phi^{10}$ potential tends to well known $\phi^6$ potential, $V^{(6)}=\frac{1}{2}\phi^2(1-\phi^2)^2$ with the asymmetric kink solution $\phi^{(6)}=\sqrt{\frac{1+\tanh(x)}{2}}$, the mass of $\frac{1}{4}$ and without any shape mode for a single kink or antikink \cite{Dorey.PRL.2011}. As the parameter $a$ increases, only up kink $\phi_{(\frac{1}{\sqrt{e^{2a}+1}},1)}^{(10)}$ remains while down kink $\phi_{(0,\frac{1}{\sqrt{e^{2a}+1}})}^{(10)}$ is eliminated.

\begin{figure*}[!ht]
\begin{center}
  \centering
    \subfigure[ ]{\includegraphics[width=0.45
 \textwidth]{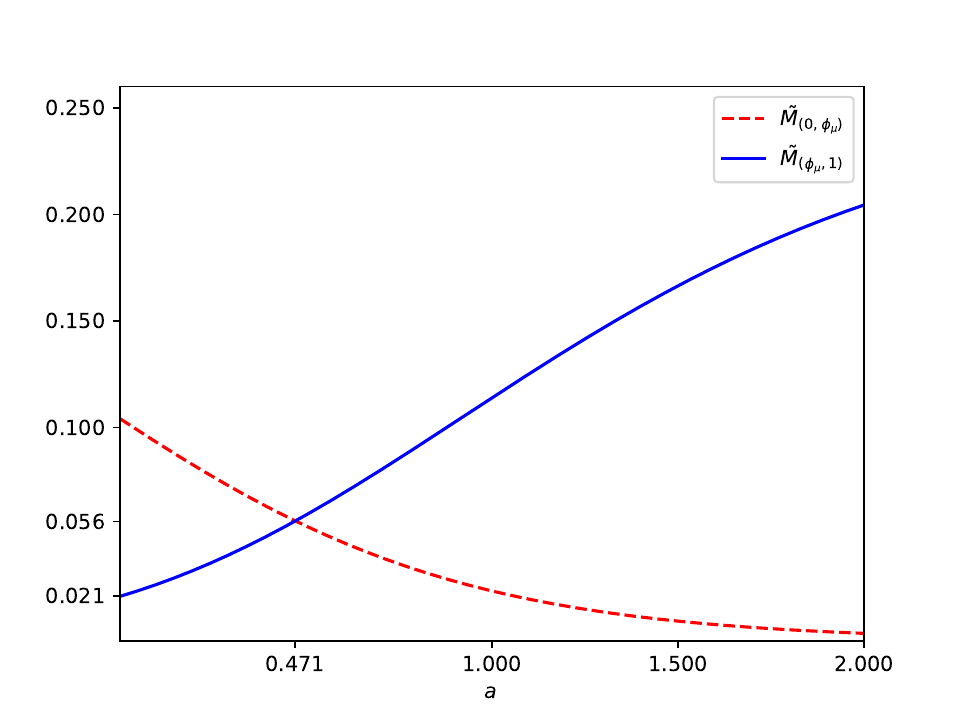}\label{fig:modifiedphi10mass}}
    \subfigure[ ]{\includegraphics[width=0.45
 \textwidth]{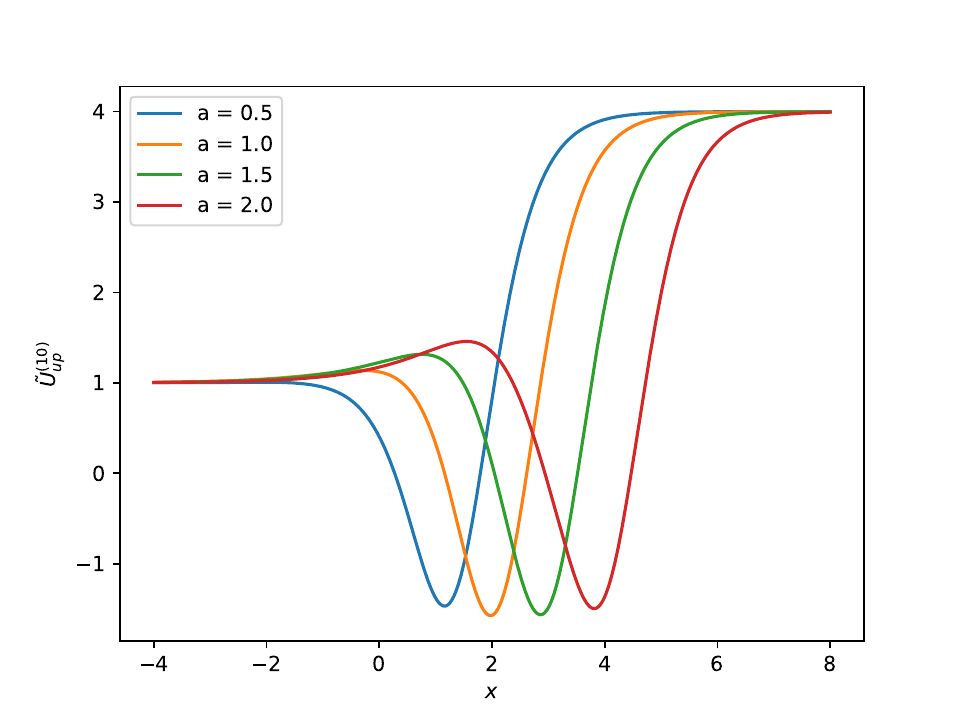}\label{fig:modifiedphi10qmpUp}}
\\
  \subfigure[ ]{\includegraphics[width=0.45
 \textwidth]{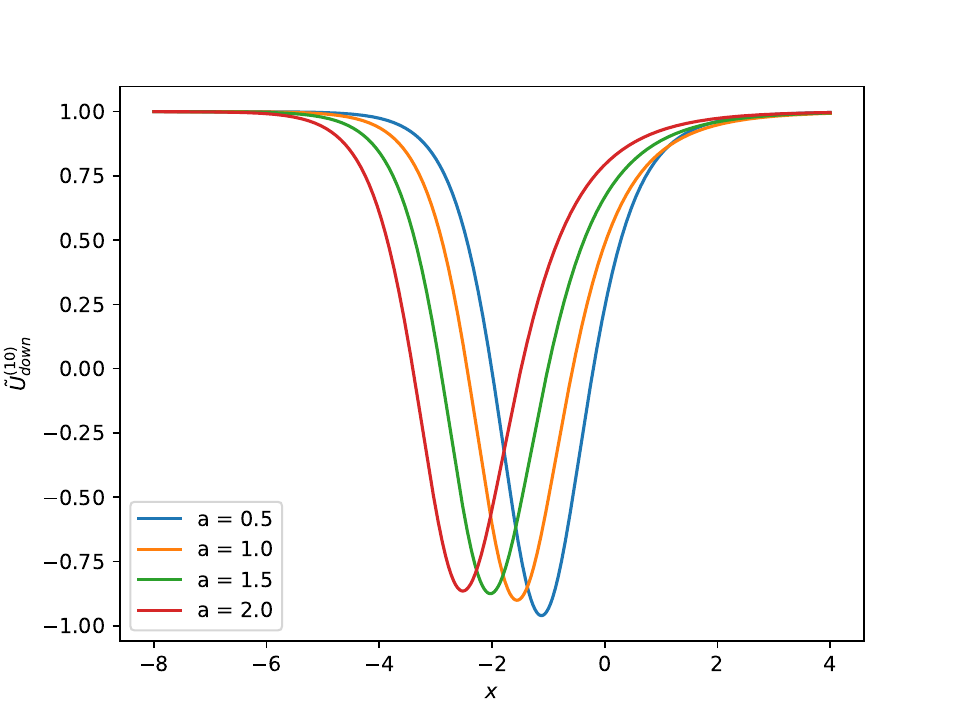}\label{fig:modifiedphi10qmpDown}}
  \subfigure[ ]{\includegraphics[width=0.45
 \textwidth]{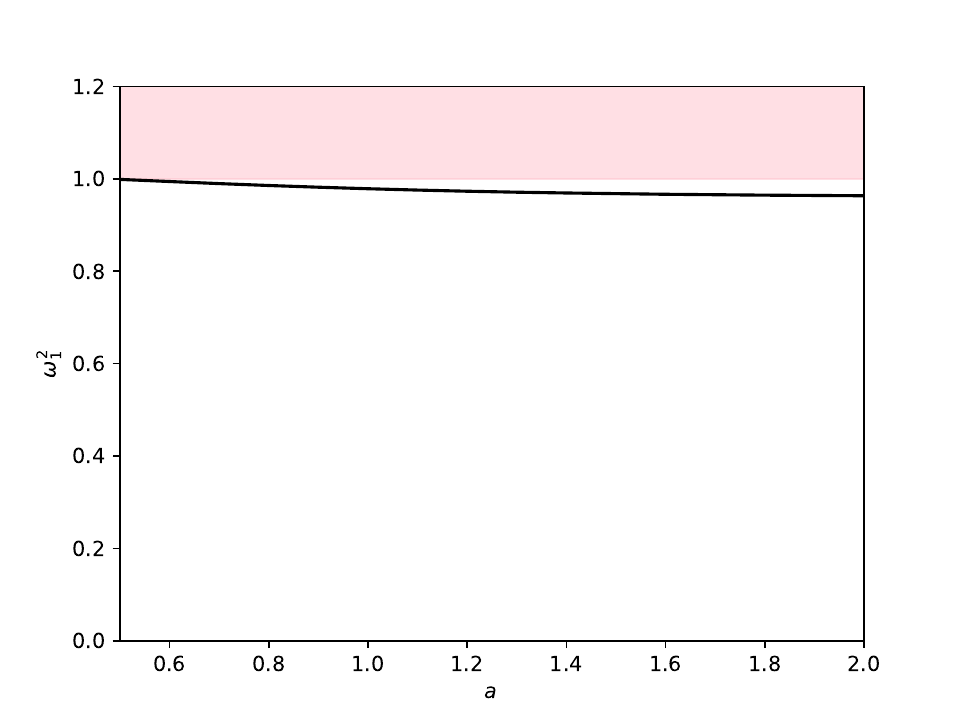}\label{fig:modifiedphi10modes}}
\\
  \caption{Deformed $\phi^{10}$ model: (a) kink mass, (b) and (c) quantum mechanical potential for up and down kinks as a function of $x$, respectively, and (d) the squared frequencies $\omega^2_1$ of the vibrational state as function of parameter $a$ for down kink.}
  \label{fig:modifiedphi10massmodesqmp}
\end{center}
\end{figure*}

The mass of these kinks are as the following, which is plotted in Fig.~\ref{fig:modifiedphi10mass}.
\begin{eqnarray}\label{eq:modifiedphi10mass}
mass &=& \int_{-\infty}^{+\infty}(\frac{\partial\phi}{\partial x})^2 dx \nonumber\\
&=&
\begin{cases}
\tilde{M}_{(0,\frac{1}{\sqrt{e^{2a}+1}})}^{(10)}= \frac{-4 e^{2 a}+e^{4 a}+2 e^{6 a}-4 e^{4 a} \left(e^{2 a}+1\right) \log \left(\frac{2 e^{2 a}}{e^{2 a}+1}\right)+1}{4 \left(\left(1-e^{2 a}\right)^3 \left(e^{2 a}+1\right)\right)},   \\
\tilde{M}_{(\frac{1}{\sqrt{e^{2a}+1}},1)}^{(10)}=\frac{1}{16} e^a \text{csch}^2(a) \left(\sinh (a)\!+\!\cosh (a)\!+\!\text{sech}(a)\!+\!2 \text{csch}(a) \ln \big(\frac{\tanh (a)+1}{e^{2 a}} \big)\right)\;\;   
\end{cases}
\end{eqnarray}
We can check that at $a \approx 0.471354$, modified $\phi^{10}$ kinks have the same mass, $mass \approx 0.0562126$.

Finally, similar to the previous section, we deal with the stability potential and we arrive at the Schr\"odinger-like equation

\begin{eqnarray}
  -\eta_{xx} + \tilde{U}^{(10)} = \omega^2 \eta   
\end{eqnarray}
where the effective potential is given by

\begin{eqnarray}\label{eq:qmpphi10}
\tilde{U}^{(10)}=\frac{d^2\tilde{V}^{(10)}(\phi)}{d\phi^2}|_{\phi=\tilde{\phi}^{(10)}(x)}.
\end{eqnarray}

These potentials are plotted in Figs.~\ref{fig:modifiedphi10qmpUp} and \ref{fig:modifiedphi10qmpDown} for up and down kink, respectively. It is important to note that the potential $\tilde{U}^{(10)}$ is not symmetrical with respect to the reflection $x \to -x$. The potential $\tilde{U}_{up}^{(10)}$ for the kink is similar to that obtained in Ref. \cite{Dorey.PRL.2011} for the $\phi^6$ model for small values of $a$. The increase in the parameter shows the appearance of a maximum in the potential. The numerical analysis of this potential only favors the appearance of the zero mode for the individual kink or antikink. However, while examining the collective antikink-kink pair, some internal modes can be found. The potential for the down kink $\tilde{U}_{down}^{(10)}$ is depicted in Fig.~\ref{fig:modifiedphi10qmpDown}. Note the presence of equal asymptotic limits and a minimum that moves away from the origin with increasing $a$. Here, we also investigated the occurrence of bound states. The results are depicted in Fig.~\ref{fig:modifiedphi10modes} and show that the value of $\omega^2_1$ approaches to the continuous mode as $a$ decreases.

Now let us investigate the kink scattering process for the $\phi^{10}$ modified model. We obtain several outcomes by varying the parameter $a$ and the initial velocity. In this sense, we solved the equation of motion with $4^{th}$ order finite-difference method with a spatial step $\delta x=0.05$. For the time dependence we used a $6^{th}$ order symplectic integrator method with a time step $\delta t=0.02$. We fixed $x_0=\pm 10$ for the initial position of the pair. 


\subsection{Down kink scattering}


We will discuss here the kink-antikink collision of the down solution $\tilde{\phi}_{(0,\frac{1}{\sqrt{e^{2a}+1}})}^{(10)}$. For the sake of simplicity, the kink is expressed as $\tilde{\phi}_{d}^K$ and the antikink as $\tilde{\phi}_{d}^{\bar{K}}$. For numerical solutions, we used the following initial conditions
\begin{eqnarray}
    \tilde\phi(x,x_0,v,0) &=& \tilde{\phi}_{d}^K(x+x_0,v,0)+\tilde{\phi}_{d}^{\bar{K}}(x-x_0,-v,0)-\phi_{\mu},\\
    \dot{\tilde \phi}(x,x_0,v,0) &=& \dot{\tilde{\phi}}_{d}^K(x+x_0,v,0)+\dot{\tilde{\phi}}^{\bar{K}}_{d}(x-x_0,-v,0),
\end{eqnarray}
where $\tilde\phi(x,x_0,v,t)=\tilde\phi(\gamma(x-vt))$ means a boost for the static solution with $\gamma=(1-v^2)^{-1/2}$ and $\phi_{\mu}=\frac{1}{\sqrt{e^{2a}+1}}$ is one vacuum of the theory. 

\begin{figure*}[!ht]
\begin{center}
  \centering
\subfigure[ ]{\includegraphics[{angle=0,width=5cm,height=4cm}]{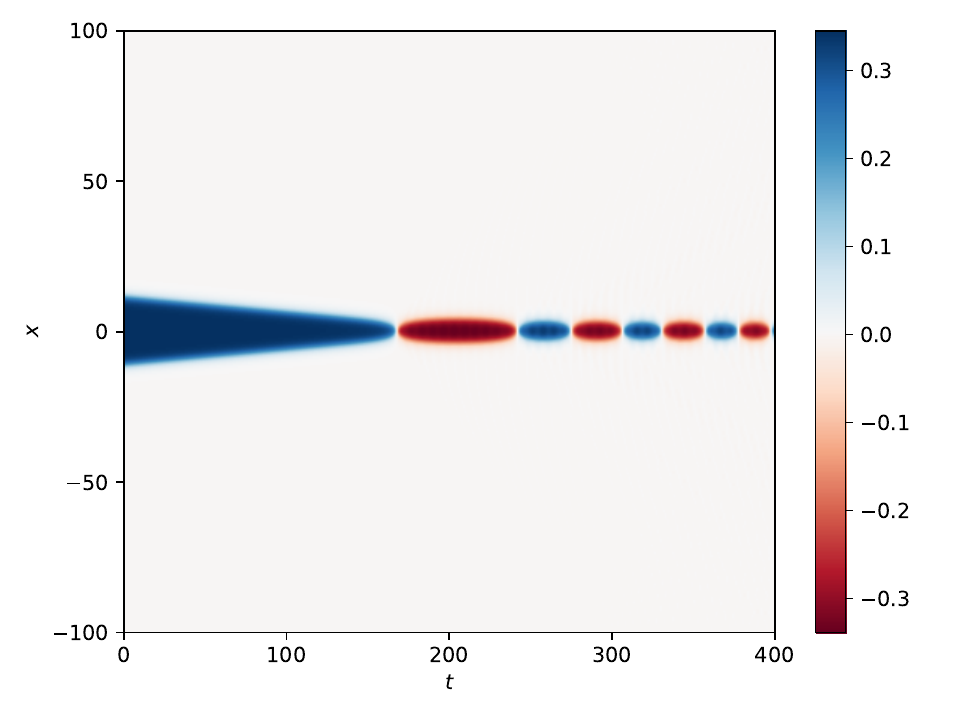}\label{ka_sd_a1_v005}}
\subfigure[ ]{\includegraphics[{angle=0,width=5cm,height=4cm}]{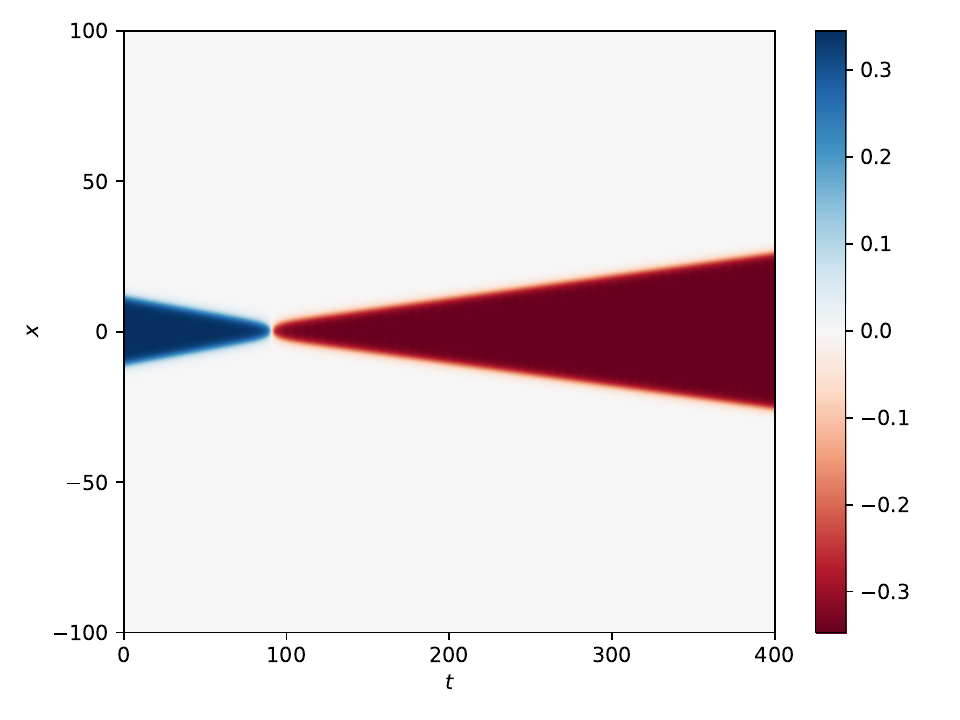}\label{ka_sd_a1_v01}}
  \caption{Kink-antikink - Evolution of scalar field $\tilde{\phi}_{d}$ in spacetime for $a=1.0$ with (a) $v=0.05$ and (b) $v=0.10$.}
\label{col_ka_sd_a1}
\end{center}
\end{figure*}

We start the analysis of collisions by varying the initial velocity $v$. The Fig. \ref{col_ka_sd_a1} depicts two forms of behavior when $a=1$. Fig. \ref{ka_sd_a1_v005} corresponds to the bion state, where the scalar field oscillates erratically at the center of mass after the interaction. The other behavior for this scattering is marked by topological sector changing. In this case, the pair approaches, collides and the scalar field exchanges vacuum - see Fig. \ref{ka_sd_a1_v01}. 

\begin{figure*}[!ht]
\begin{center}
  \centering
\subfigure[ ]{\includegraphics[{angle=0,width=5cm,height=4cm}]{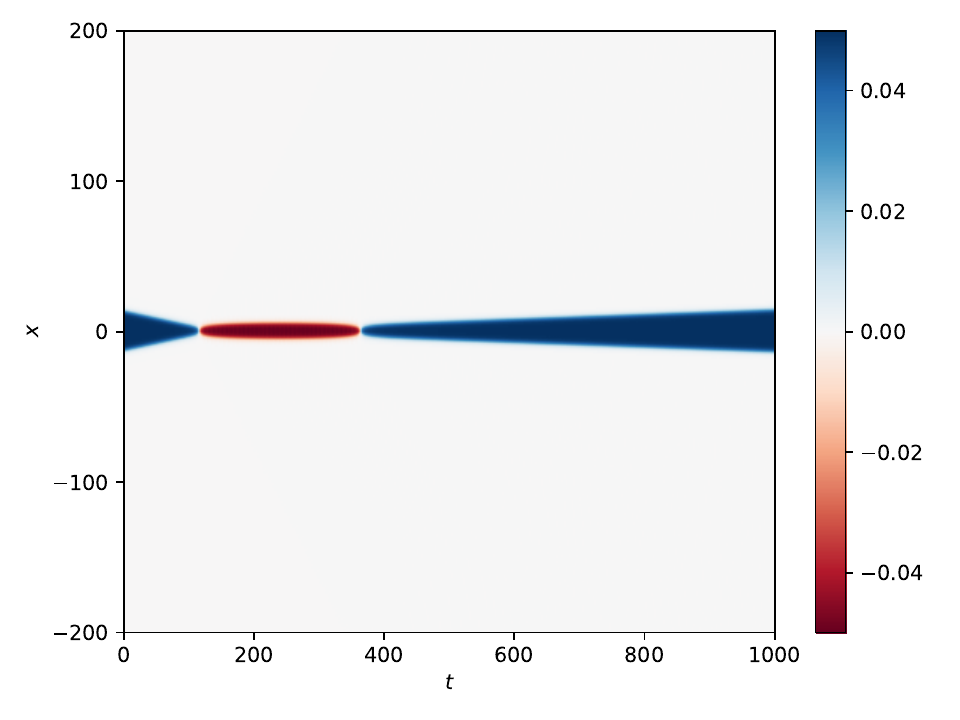}\label{ka_sd_a3_v00934}}
\subfigure[ ]{\includegraphics[{angle=0,width=5cm,height=4cm}]{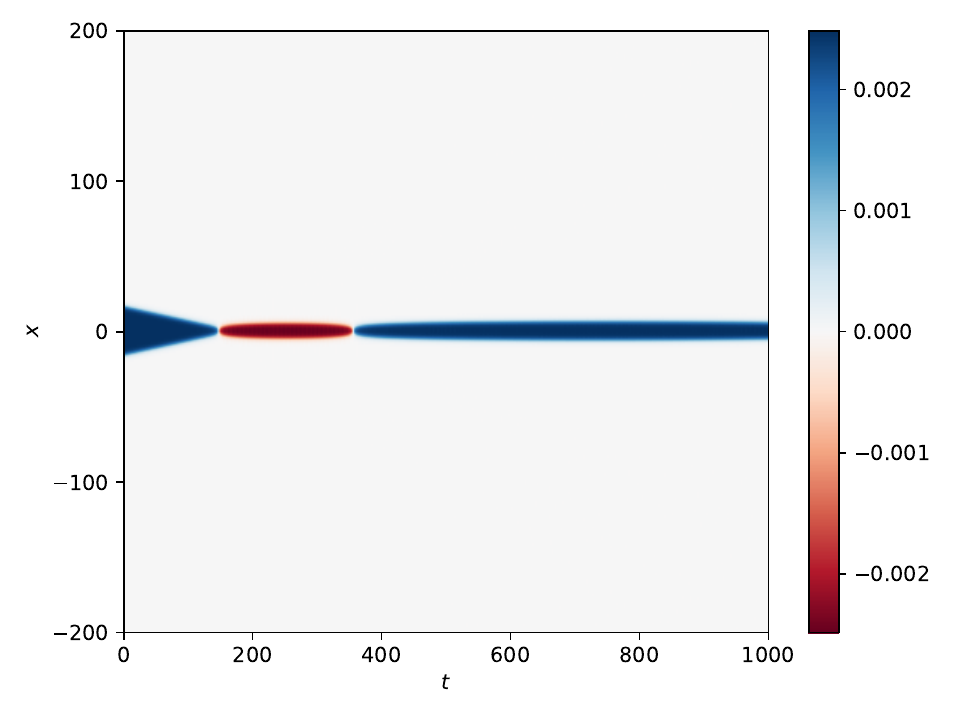}\label{ka_sd_a6_v00936}}
  \caption{Kink-antikink - Evolution of scalar field $\tilde{\phi}_{d}$ in spacetime for (a) $a=3.0$ with $v=0.0934$ and (b) $a=6.0$ with $v=0.0936$.}
\label{col_ka_sd_a36}
\end{center}
\end{figure*}

Additionally, we altered the value of the parameter and performed exhaustive collisions for various initial velocity values. The scattering of this configuration showed the appearance of few two-bounce behaviors, indicating that the two-bounce windows for this case are very sensitive. Note the need for a high value of the final time in order to observe scattering completely. For instance, in Fig. \ref{col_ka_sd_a36} we note that the kink-antikink pair approaches, collides, and after a long period, collides again and then entirely moves away. Notice that the value of the field decreases as the value of $a$ increases, compare Fig. \ref{ka_sd_a3_v00934} for $a=3$ with Fig. \ref{ka_sd_a6_v00936} for $a=6$.

We shall now explore the antikink-kink scattering of the solution $\tilde{\phi}^K_d$. The initial conditions used in this problem are given by
\begin{eqnarray}
    \tilde\phi(x,x_0,v,0) &=& \tilde{\phi}_{d}^{\bar K}(x-x_0,-v,0)+\tilde{\phi}_{d}^{K}(x+x_0,v,0),\\
    \dot{\tilde \phi}(x,x_0,v,0) &=& \dot{\tilde{\phi}}_{d}^{\bar K}(x-x_0,-v,0)+\dot{\tilde{\phi}}^{K}_{d}(x+x_0,v,0),
\end{eqnarray}
where $\tilde\phi(x,x_0,v,t)=\tilde\phi(\gamma(x-vt))$ means a boost for the static solution with $\gamma=(1-v^2)^{-1/2}$. In this scenario, two-bounce windows are absent for small values of $a$. However, the development of the resonant structure is facilitated by the increase of this parameter.

\begin{figure*}[!ht]
\begin{center}
  \centering
\subfigure[ ]{\includegraphics[{angle=0,width=5cm,height=4cm}]{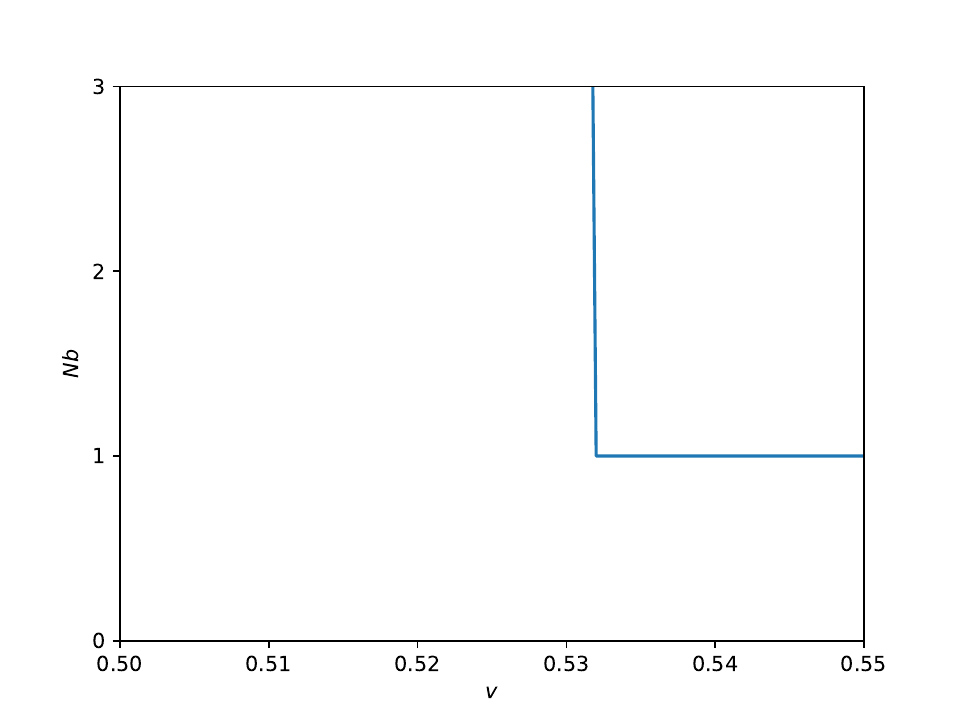}\label{ak_Nbxv_sda1}}
\subfigure[ ]{\includegraphics[{angle=0,width=5cm,height=4cm}]{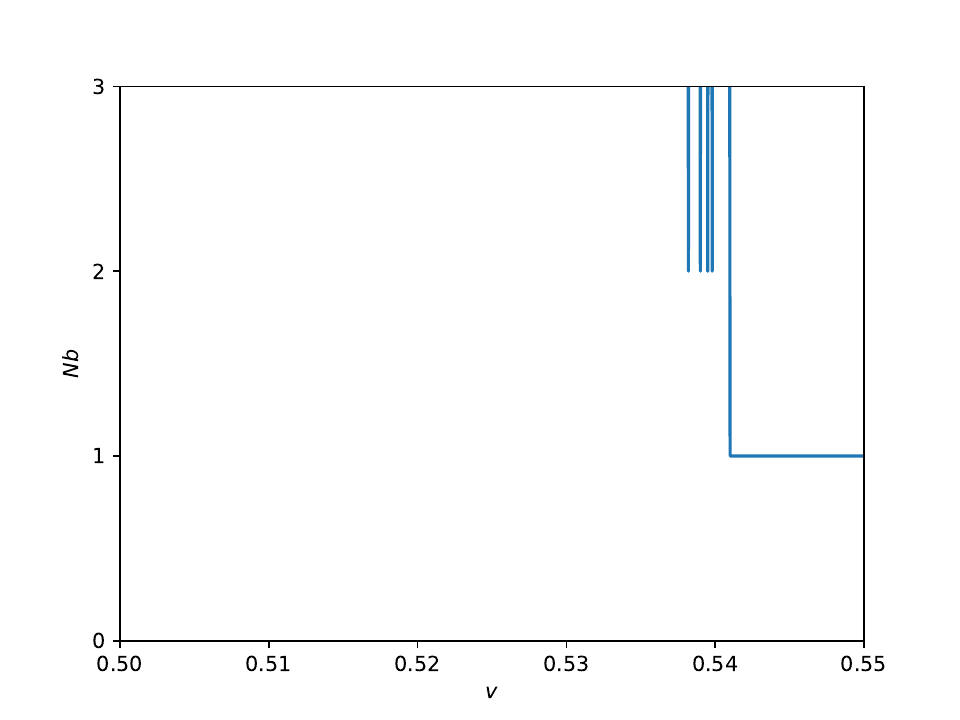}\label{ak_Nbxv_sda3}}
\subfigure[ ]{\includegraphics[{angle=0,width=5cm,height=4cm}]{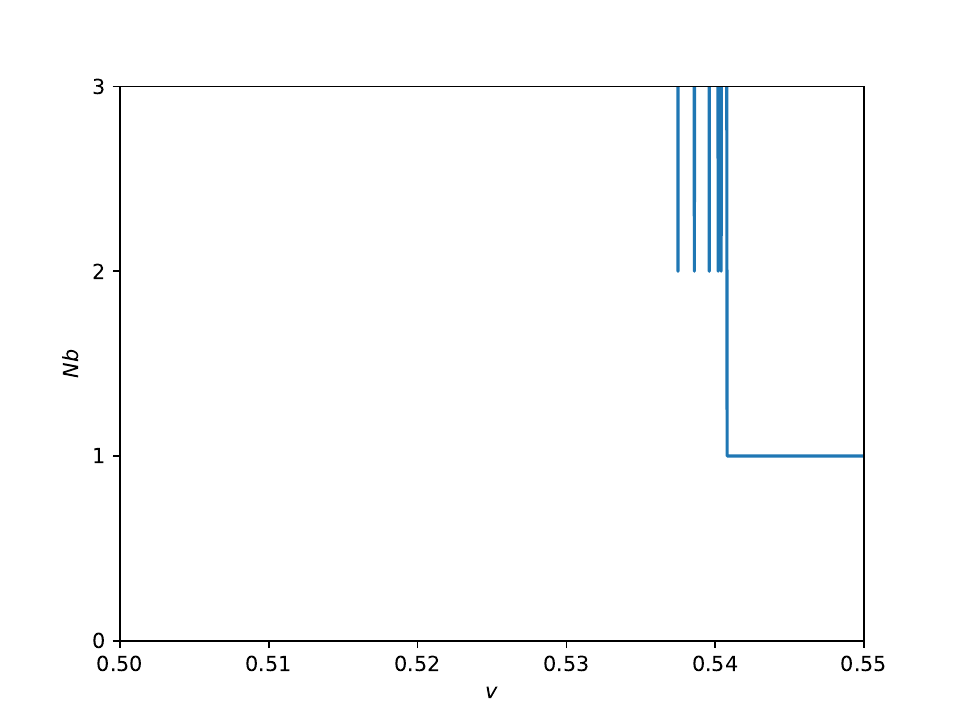}\label{ak_Nbxv_sda6}}
  \caption{Antikink-kink - Number of bounces versus initial velocity of scalar field $\tilde{\phi}_{(0,\frac{1}{\sqrt{e^{2a}+1}})}^{(10)}$ for (a) $a=1$, (b) $a=3$ and (c) $a=6$.}
\label{ak_Nbxv_sd_a}
\end{center}
\end{figure*}

The number of bounces relative to the initial velocity is computed in Fig. \ref{ak_Nbxv_sd_a}. Take note that there is no two-bounce window for $a=1$ Figs \ref{ak_Nbxv_sda1}. Only two regions are created for this value of $a$, the region with inelastic collisions for velocities greater than the critical velocity and the region with bion states for $v<v_c=0.532$. In contrast, as illustrated in Figs. \ref{ak_Nbxv_sda3} and \ref{ak_Nbxv_sda6}, double collisions happen for greater values of $a$. We can see that at certain velocity levels, the kinks collide twice and then separate. The behavior with two collisions is denoted by $Nb=2$, whereas the behavior with only one collision is marked by $Nb=1$. The value of the critical velocity increases as a result of increasing the parameter.

The proximity of the vibrational mode value to the continuous spectrum serves as the basis for the explanation of why the two-bounce windows vanish for small values of $a$. The continuous mode in this configuration is obtained for frequencies higher than $\omega^2>1$. Notice that the value of $\omega_1^2$ gets closer and closer to $1$ as $a$ decreases. As a result of the elimination of the bound mode in the continuum, a quasinormal mode forms. This is a necessary component of the phenomenon known as spectral walls \cite{adam.2019}.

Again, the scattering results are affected by how the initial condition is established. The kink-antikink configuration, for example, has its asymptotic values invariant, however, the center of the collision shows the field value decreasing as $a$ increases. In contrast, the antikink-kink scattering has its center of mass invariant and the variation of $a$ causes the asymptotic values of the defect to change.


\subsection{Up kink scattering}


In this section we will study the kink-antikink and antikink-kink collision of the up solution $\tilde{\phi}_{(\frac{1}{\sqrt{e^{2a}+1}},1)}^{(10)}$. For the sake of simplicity, the kink is written as $\tilde{\phi}_{u}^K$ and the antikink as $\tilde{\phi}_{u}^{\bar{K}}$. In this first part, the kink-antikink scattering we will be examined. We used the following initial conditions
\begin{eqnarray}
    \tilde\phi(x,x_0,v,0) &=& \tilde{\phi}_{u}^K(x+x_0,v,0)+\tilde{\phi}_{u}^{\bar{K}}(x-x_0,-v,0)-1,\\
    \dot{\tilde \phi}(x,x_0,v,0) &=& \dot{\tilde{\phi}}_{u}^K(x+x_0,v,0)+\dot{\tilde{\phi}}^{\bar{K}}_{u}(x-x_0,-v,0),
\end{eqnarray}
where $\tilde\phi(x,x_0,v,t)=\tilde\phi(\gamma(x-vt))$ means a boost for the static solution with $\gamma=(1-v^2)^{-1/2}$. It is important to note here that there is no vibrational mode in this configuration, even for the kink-antikink pair. A center barrier is produced by the perturbation potential for the collective kink-antikink pair, making the presence of bound states impossible. This behavior is similar to the potential in Ref. \cite{Dorey.PRL.2011}. As a consequence, the two-bounce windows are absent.

\begin{figure*}[!ht]
\begin{center}
  \centering
\subfigure[ ]{\includegraphics[{angle=0,width=4cm,height=4cm}]{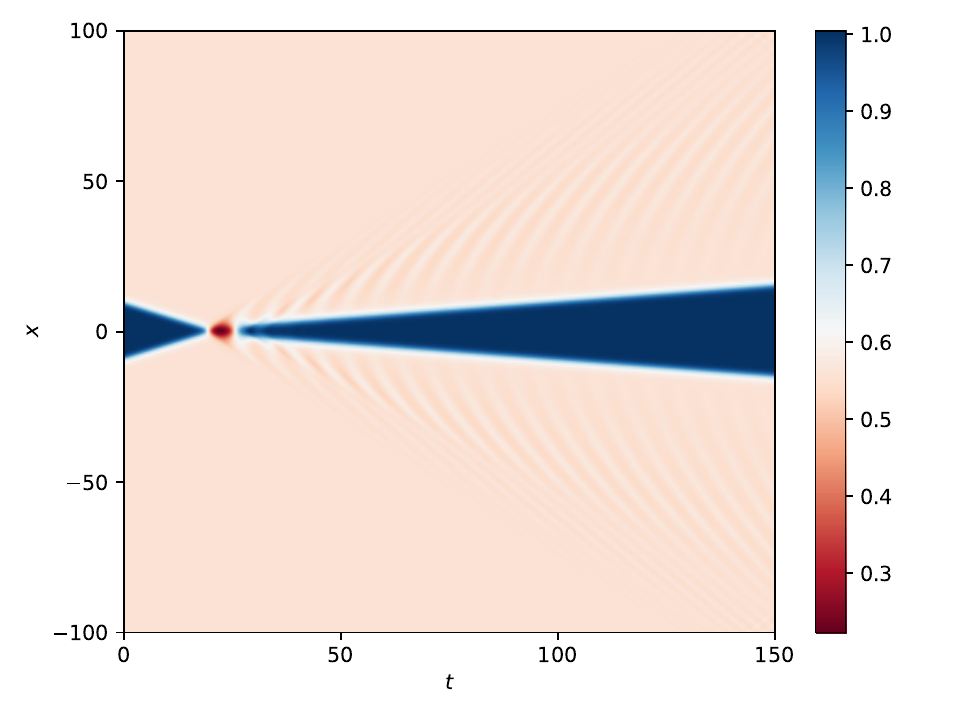}\label{ka_su_a04_v045}}
\subfigure[ ]{\includegraphics[{angle=0,width=4cm,height=4cm}]{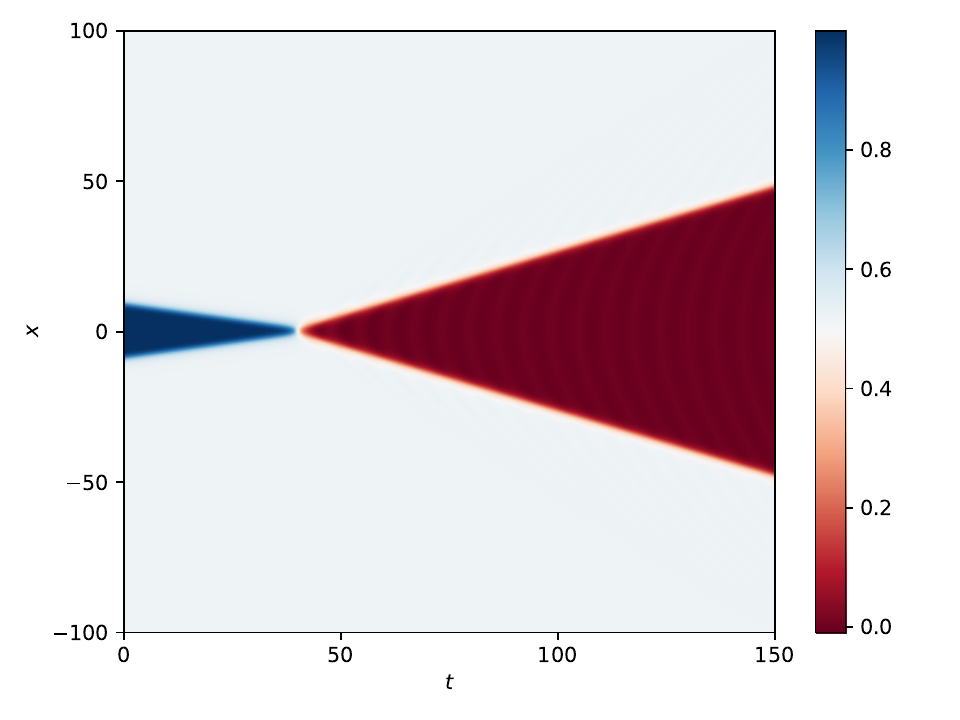}\label{ka_su_a05_v02}}
\subfigure[ ]{\includegraphics[{angle=0,width=4cm,height=4cm}]{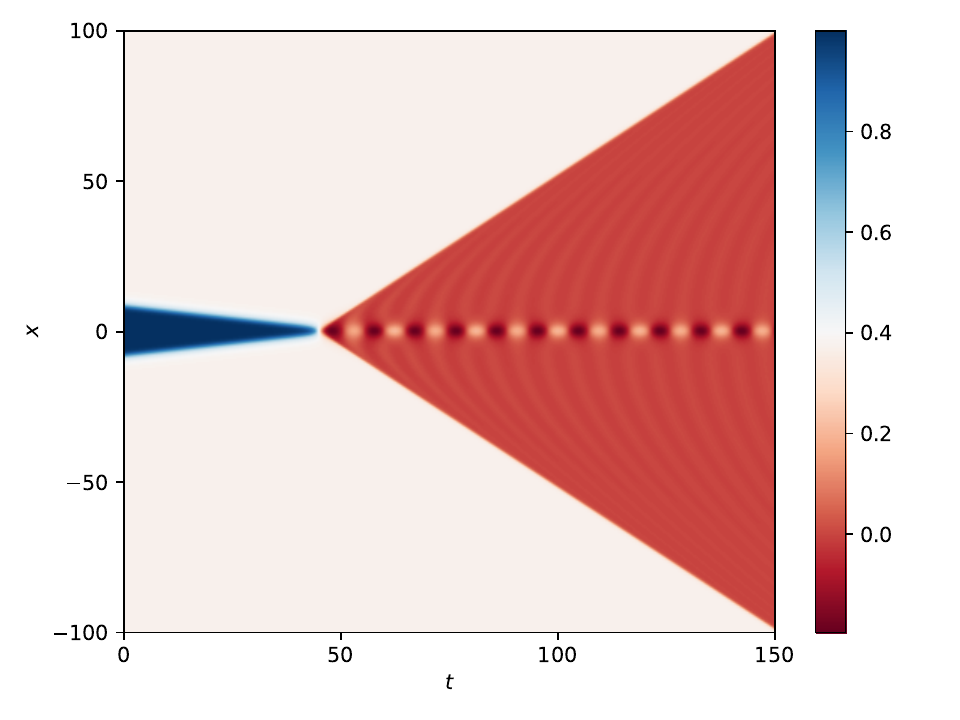}\label{ka_su_a09_v016}}
\subfigure[ ]{\includegraphics[{angle=0,width=4cm,height=4cm}]{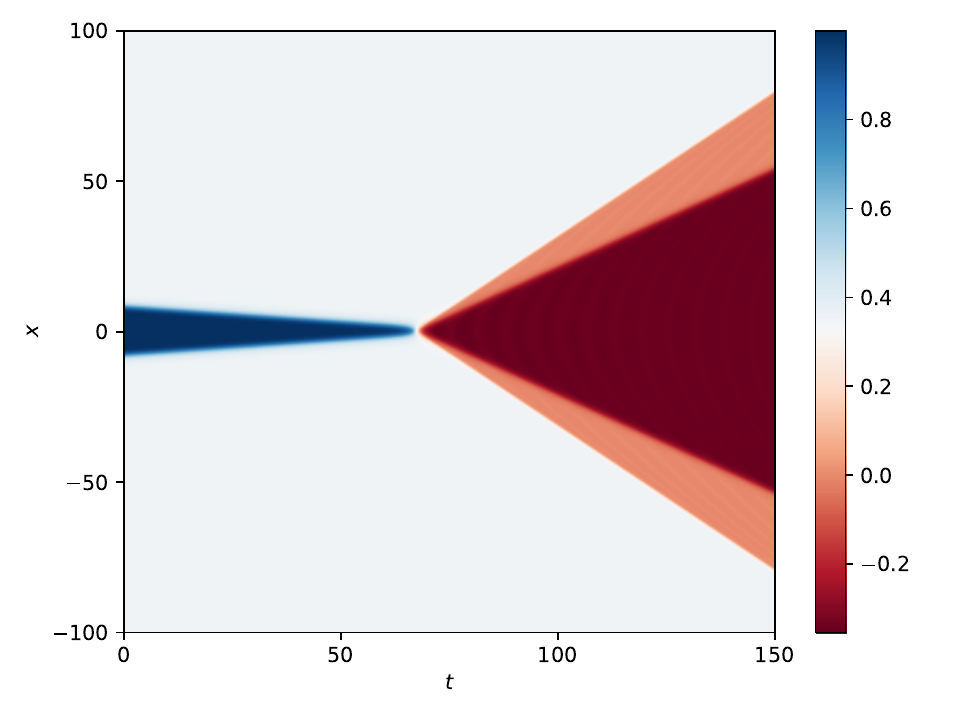}\label{ka_su_a1_v01}}
  \caption{Kink-antikink - Evolution of scalar field $\tilde{\phi}_{u}$ in spacetime for (a) $a=0.4$ with $v=0.45$ (b) $a=0.5$ with $v=0.2$, (c) $a=0.9$ with $v=0.16$ and (d) $a=1.0$ with $v=0.1$.}
\label{col_ka_su_a}
\end{center}
\end{figure*}

Some results are achieved when we vary the values of the initial velocity and the parameter $a$. For instance, the kink-antikink pair approaches, collides, and eventually separates from one another in Fig. \ref{ka_su_a04_v045} when $a=0.4$. This result occurs for velocities above the critical velocity $v>v_c=0.4332$. On the other hand, when the collision occurs for small values of the initial velocity, we observe only a bion behavior. The behavior of the scattering begins to change when $a>0.5$. In this case, we realize only the formation of an antikink-kink down pair after the collision, as shown in Fig. \ref{ka_su_a05_v02}. It should be noted that the velocity after the collision is greater than the approach velocity of the kink-antikink pair. This is due to the up kink being more massive than the down kink. Consequently, there is kinetic energy to accelerate the newly formed anti-kink-kink pair. Additionally, the increase in $a$ causes an oscillation at the center of mass - see Fig. \ref{ka_su_a09_v016}. This behavior favors the perturbation of the pair and consequently, we note the presence of a double antikink-kink Fig. \ref{ka_su_a1_v01}. Notice that the kinks are now part of the other topological sector. Another feature is that the new pair scatters with less radiation. These behaviors are comparable to those observed in reference \cite{simas.2020}.

Let us now examine the antikink-kink scattering process of the up solution $\tilde{\phi}_{(\frac{1}{\sqrt{e^{2a}+1}},1)}^{(10)}$. The initial conditions for this configuration are expressed as follows
\begin{eqnarray}
    \tilde\phi(x,x_0,v,0) &=& \tilde{\phi}_{u}^{\bar K}(x-x_0,-v,0)+\tilde{\phi}_{u}^K(x+x_0,v,0)-\phi_{\mu},\\
    \dot{\tilde \phi}(x,x_0,v,0) &=& \dot{\tilde{\phi}}_{u}^{\bar K}(x-x_0,-v,0)+\dot{\tilde{\phi}}^K_{u}(x+x_0,v,0),
\end{eqnarray}
where $\tilde\phi(x,x_0,v,t)=\tilde\phi(\gamma(x-vt))$ means a boost for the static solution with $\gamma=(1-v^2)^{-1/2}$ and $\phi_{\mu}=\frac{1}{\sqrt{e^{2a}+1}}$. Once again, an excited mode is not revealed by the perturbation analysis for a single kink or antikink. However, the Schr\"odinger potential ensures the existence of numerous bound states when we analyze the collective antikink-kink pair. This is favored due to the configuration of how the initial condition is written, generating a perturbation potential with a central well and asymptotic to $\tilde{U}^{(10)}_{up} \to 4$ with $x\to \pm \infty$. The Ref. \cite{Dorey.PRL.2011} also exhibits this feature.

\begin{figure*}[!ht]
\begin{center}
  \centering
\subfigure[ ]{\includegraphics[{angle=0,width=5cm,height=4cm}]{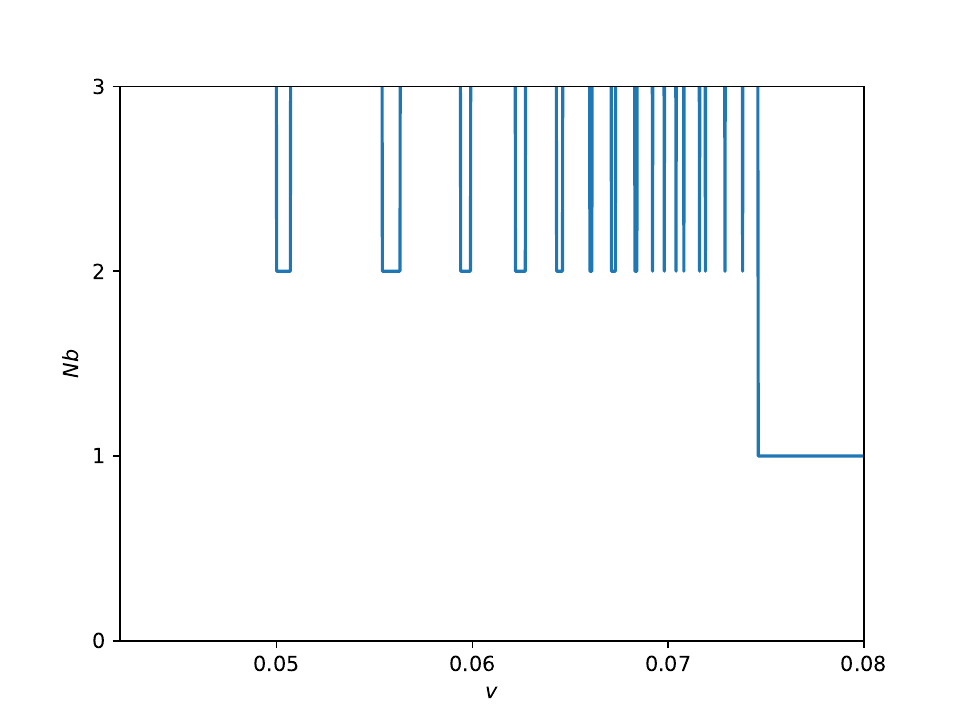}\label{ak_Nbxv_sua01}}
\subfigure[ ]{\includegraphics[{angle=0,width=5cm,height=4cm}]{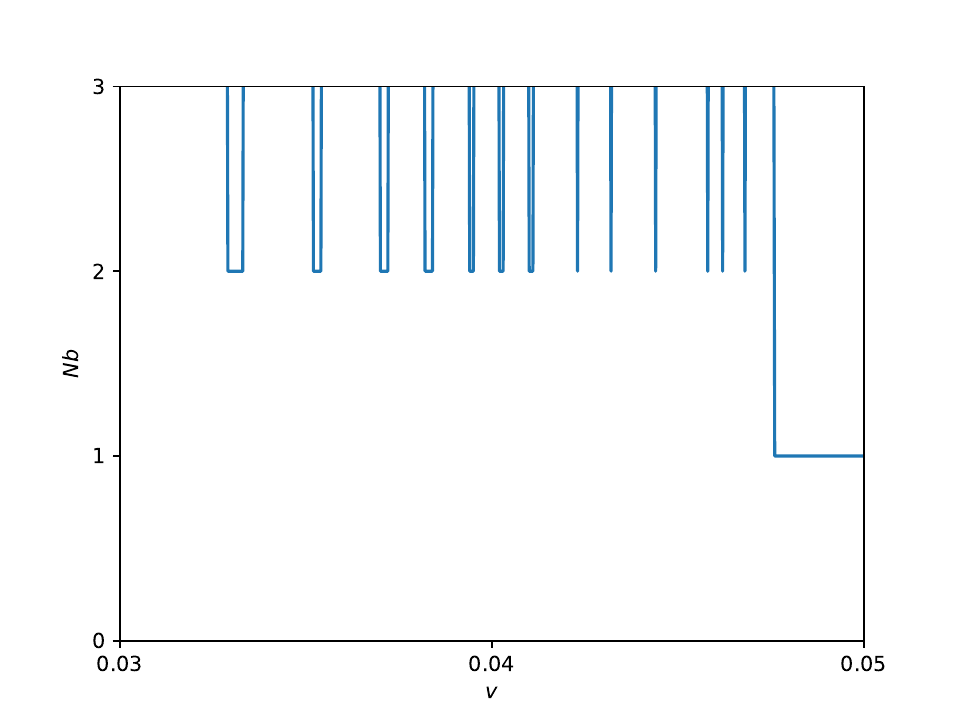}\label{ak_Nbxv_sua03}}
  \caption{Antikink-kink - Number of bounces versus initial velocity of scalar field $\tilde{\phi}_{(\frac{1}{\sqrt{e^{2a}+1}},1)}^{(10)}$ for (a) $a=0.1$ and (b) $a=0.3$.}
\label{ak_Nbxv_su_a}
\end{center}
\end{figure*}

The presence of vibrational modes therefore denotes the emergence of two-bounce windows. However, due to the complexity of the resonant energy exchange between the extra bound modes, such resonance windows are extremely thin. The Fig. \ref{ak_Nbxv_su_a} depicts the structure of antikink-kink scattering, which includes the number of collisions relative to the initial velocity. The result for $a=0.1$ is shown in Fig. \ref{ak_Nbxv_sua01}. For $v>v_c=0.0746$, it describes the region with just one collision ($Nb=1$). For velocities smaller than the critical velocity, we observe the bounces structure between the bion region and with more collisions. One can see that the critical velocity decreases with increasing parameter $a$. This can be seen in Fig. \ref{ak_Nbxv_sua03} for $a=0.3$, where $v_c=0.0476$. The critical velocity continues decreasing as $a$ increases, for instance, for $a=1.0$, we realize that $v_c=0.0379$. This illustrates the zone of bion states contracting and the region of inelastic collision expanding.


\section{Conclusion}
\label{sec:conclusion}


In this work we have used the deformation procedure to introduce two new models, the modified $\phi^8$ and  $\phi^{10}$ models. These models are respectively obtained by deforming models $\phi^4$ and $\phi^6$ with the aid of deformation techniques. The models contain a real parameter $a$. We have shown that this parameter controls the potentials, the kinklike solutions, their masses, the stability potentials and the internal modes associated to the stability of the systems. 

The modified $\phi^8$ model contains both types of symmetric and asymmetric kinks. The stability analysis showed that the asymmetric kink has a shape mode, while the symmetric kink has an internal mode for $a \geq 4.6$. We performed the kink collision for both solutions. We investigated the kink-antikink and antikink-kink collision processes in the asymmetric case. For small values of $a$, we observe two-bounce windows for kink-antikink. However, increasing the parameter starts the process of suppressing the resonant windows and the formation of new antikink-kink pairs. This behavior is related to the massive character of the asymmetric kink, allowing sector exchange. When we consider the antikink-kink interaction, we also see the presence of two-bounce windows for small values of $a$. The increase in this parameter shows the appearance of false two-bounce windows, until the complete annihilation of the resonances. The significance of the parameter in the initial configuration accounts for the difference in the results between kink-antikink and antikink-kink scattering. The increase in $a$ favors the appearance of a new vibrational mode for the symmetric kink, which in turn influences the internal mode of the asymmetric kink. Consequently, the resonant energy exchange mechanism between the modes is frustrated. Now, the scattering analysis for the symmetrical kink revealed the absence of fractal structures for small values of $a$. However, the formation of antikink-kink pairs in the asymmetrical sector was noticeable. Furthermore, large values of the parameter modify the results, indicating the presence of a resonant structure. In addition, with $a \to \infty$, the deformed $\phi^8$ tends towards the $\phi^4$ model.

The $\phi^{10}$ model contains five minima and two forms of asymmetric kinks known as down and up kink. Considering small perturbations around the static solutions, we arrive at a Schr\"odinger equation with different perturbation potentials for the down and up kink. We numerically investigated the occurrence of vibrational modes at both potentials. However, we observed the presence of only one shape mode for the potential for an individual down kink. Numerical analysis of kink-antikink and antikink-kink scattering showed complete suppression of the resonant windows for small values of $a$. However, as this parameter increased, some two-bounce behavior was observed. The proximity of the square frequency to the continuous spectrum serves as the basis for explaining why the two-bounce windows disappear for small values of $a$. Furthermore, we investigated the kink-antikink collision for the up kink. The results in this scenario reveal the emergence of a more accelerated antikink-kink down pair after the collision. This is due to the up kink being more massive than the down kink. Resonant windows are absent in this case because the perturbation potential of the kink-antikink pair generates a barrier, which inhibits the presence of the bound modes. Finally, antikink-kink scattering indicates the appearance of two-bounce windows. We can notice that the excited modes for this case are revealed by the perturbation analysis for the antikink-kink pair.

One of the issues that can arise when researching models with a high-order of self-interaction is kink scattering. Specifically, there is the possibility of different initial conditions because of the number of minima in the potential of the $\phi^8$ and $\phi^{10}$ models, which results in a multitude of effects after the collision. For instance, the scattering outputs that appear in the higher-order models go beyond those found in the double sine-Gordon model and the $\phi^4$ theory. Thus, as models with high-order self-interaction  may be related to the understanding of successive first-order phase transitions, further research is still required to obtain a better knowledge of kink-antikink and antikink-kink scattering for these models.

We would also like to mention some issues that may be of interest for future studies. This includes, for example, the investigation of the force between the kinks using the Manton method \cite{manton2}. Additionally, explicit solutions for higher-order models can be obtained by introducing new deformation functions. The study of the scattering of these solutions, in particular, can create new scenarios for the kink-antikink and antikink-kink collisions to enlarge the relationship between vibrational and translational modes.

Unlike what happens in the $\phi^4$ model, where the tail is exponential, some higher-order models can exhibit topological kinks with power-law tails, providing long-range interactions for the solutions. Therefore, another perspective for future research is to investigate in detail the scattering structure that can be produced for the scenario $a=1$, which exhibits a kink with a long-range tail. Remarkably, Rydberg atoms exhibit a long-range interaction, because they interact as a dipole-dipole when an electron is excited to an energy level with a very high value of the primary quantum number $n$ \cite{saffman}. Applications in the realm of quantum information are now made possible by the strong long-range interaction between Rydberg atoms.

Similar to what occurs in the $\phi^6$ model \cite{Dorey.PRL.2011}, certain $\phi^8$ models that have been explored explain that the development of resonant windows is associated with the mechanism of resonant exchange between the bound states of the collective kink-antikink pair \cite{EkaGani}. In the present work, we noticed the presence of a set of internal modes, rather than only the particular kink or kink-antikink combination. Related to the presence of the vibrational mode, it can be useful to further development the collective coordinates method to better understand interactions in higher-order models.


\section*{Acknowledgements}
AMM would like to thank Islamic Azad University Quchan branch for the grant. This study was financed in part by the Coordena\c c\~ao de Aperfei\c coamento de
Pessoal de N\'ivel Superior - Brasil (CAPES) - Finance Code 001. It was also financed in part by Conselho Nacional de Desenvolvimento Cient\'ifico e Tecnol\'ogico, Grant No. 303469/2019-6 (DB) and by Paraiba State Research Foundation, Grant No. 0015/2019 (DB).


\end{document}